\pdfoutput=1
\documentclass[a4paper,fleqn,usenatbib,longnamesfirts]{mnras}
\usepackage[T1]{fontenc}
\usepackage{ae,aecompl}

\usepackage{graphicx}	% Including figure files
\usepackage{amsmath}	% Advanced maths commands
\usepackage{amssymb}	% Extra maths symbols
\usepackage{subfig}
\usepackage{multirow}
\usepackage{color}
\usepackage{etoolbox}
\makeatletter
\patchcmd\@combinedblfloats{\box\@outputbox}{\unvbox\@outputbox}{}{%
	\errmessage{\noexpand\@combinedblfloats could not be patched}%
}%
\makeatother

\title[Morphological analysis of MUSIC clusters SZ maps]{Morphological estimators on Sunyaev--Zel'dovich maps of MUSIC clusters of galaxies}

\author[G. Cialone et al.]{%
Giammarco Cialone,$^{1}$\thanks{E-mail: giammarco.cialone@uniroma1.it}
Marco De Petris,$^{1}$
Federico Sembolini,$^{1,2,3}$
Gustavo Yepes,$^{2,3}$
\newline
\newauthor Anna Silvia Baldi,$^{1,4}$ and Elena Rasia$^{5}$
\\
\\
$^{1}$Dipartimento di Fisica, Sapienza Universit\`{a} di Roma, p.le Aldo Moro 5, I-00185 Roma, Italy\\
$^{2}$Departamento de F\'{i}sica Te\'{o}rica, M\'{o}dulo 8, Facultad de Ciencias, Universidad Aut\'{o}noma de Madrid, E-28049 Cantoblanco, Madrid, Spain\\
$^{3}$Astro-UAM, UAM, Unidad Asociada CSIC, E-28006, Madrid, Spain\\
$^{4}$Dipartimento di Fisica, Universit\`{a} di Roma Tor Vergata, via della Ricerca Scientifica 1, I-00133 Roma, Italy\\
$^{5}$INAF, Osservatorio Astronomico di Trieste, via Tiepolo 11, I-34131, Trieste, Italy\\
}

\date{Accepted XXX. Received YYY; in original form ZZZ}

\pubyear{2018}

\begin{document}
\label{firstpage}
\pagerange{\pageref{firstpage}--\pageref{lastpage}}
\maketitle

\begin{abstract}
        % MODIFIED BY E.R.
	% introduction 
The determination of the morphology of galaxy clusters has important
repercussion on their cosmological and astrophysical studies.
	% methods
In this paper we address the morphological characterisation of
synthetic maps of the Sunyaev--Zel'dovich (SZ) effect produced for a
sample of 258 massive clusters
($M_{vir}>5\times10^{14}h^{-1}$M$_\odot$ at $z=0$), extracted from the
MUSIC hydrodynamical simulations.  Specifically, we apply five known
morphological parameters, already used in X-ray, two newly introduced
ones, and we combine them together in a single parameter.  We analyse
two sets of simulations obtained with different prescriptions of the
gas physics (non radiative and with cooling, star formation and
stellar feedback) at four redshifts between 0.43 and 0.82.
	% results and discussion 
For each parameter we test its stability and efficiency to
discriminate the true cluster dynamical state, measured by theoretical
indicators.
%	We compare our results with the available X-ray counterpart for the same data set, finding a good agreement
%	for the light concentration parameter only.
	%Furthermore we introduce another parameter that properly combines the single ones. 
The combined parameter discriminates more efficiently relaxed and
disturbed clusters. This parameter had a mild correlation with the
hydrostatic mass ($\sim 0.3$) and a strong correlation ($\sim 0.8$)
with the offset between the SZ centroid and the cluster centre of
mass. The latter quantity results as the most accessible and efficient
indicator of the dynamical state for SZ studies.
\end{abstract}


%%%%%%%%%%%%%%%%%%%%%%%%%%%%%%%%%%%%%%%%%%%%%%%%%%

%%%%%%%%%%%%%%%%% BODY OF PAPER %%%%%%%%%%%%%%%%%%

\section{Introduction}
% MODIFIED BY E.R. 
% INTRODUCE SOME REFERENCES -LAU+2009, RASIA+2012, MANTZ
% INTRODUCE BETTER THE NEW PARAMETER FOR THE LATEST CORRELATION
%
\label{sec:introduction}
Clusters of galaxies are the largest virialized objects in the universe and
they play a key role in the understanding of the formation, the growth, and the properties of large-scale structures.
The investigation of cluster dynamical state has important astrophysical and
cosmological implications. The disturbed nature of a cluster can be in fact the indicator of violent mergers.
A sample of disturbed clusters would allow to characterize the motions arising from these events, which are of crucial 
importance for an accurate estimate of the non-thermal pressure support which will need to be constrained to enable
accurate cluster cosmology
\citep[see e.g.][] {Haiman_2001_constrains_on_cosmological_parameters_from_galaxy_clusters,
	Borgani_2008_Cosmology_with_clusters_of_galaxies, Mantz_2010_cosmological_parameters_from_clusters}.
As shown from numerical simulations, values of cluster masses derived only from the
assumption of hydrostatic equilibrium, and thus accounting only for the contribution from thermal pressure,
can be significantly understimated (up to $ \sim 30$ per cent)
\citep[see e.g.][and references therein]{lau.etal.2009,rasia.etal.2012,Biffi_et_al_2016:On_the_Nature_of_Hydrostatic_Equilibrium_in_Galaxy_Clusters}.
Such mass bias is reduced to 10 per cent in relaxed clusters.
This is one of the main reasons of the common choice to calibrate the self-similar scaling relations
\citep[see e.g.][for a review]{giodini:scalingrelations} towards regular objects
\citep[see e.g.][]{Johnston_2007_YM_relation_calibration,George_2012_YM_calibration,Czakon_2015_YM_calibration}.

Historically, the very first attempts to determine whether a galaxy cluster was dynamically relaxed or not,
were made through the visual inspection of the galaxy distribution in optical images \citep{Conselice2003:fluctuationparameter}
and the regularity of the intra-cluster medium (ICM) in X-ray maps 
\citep{Ulmer_1982:Three_rich_clusters_of_galaxies_with_bimodal_or_clumpy_X-ray_morphologies}.
The presence of bimodal or clumpy surface brightness distribution was considered an indication for the presence of
sub-structures in dynamically active clusters \citep{Jones_et_al_1992:Clusters_ad_superclusters_of_galaxies,
	Slezak_1994:clusters_x_ray_sub_structures, Gomez_1997:x_ray_clusters_sub_structures_observations,
	Rizza_1998:x_ray_sub_structures,Richstone_1992:sub_structures_implies_disturbed_state}, although some exceptions have been reported
\citep[e.g.][]{Pinkney_1996_substructures_in_clusters_of_galaxies,Buote_1996:morphology_and_dynamical_state,
	Plionis_2002_clusters_dynamical_evolution}.
On the other hand, regular and mostly circular shapes of the projected gas distribution are characteristic of dynamically
relaxed clusters, which have limited turbulent motions.
From these early studies, several other morphological parameters have been introduced
in the literature over the years (see section~\ref{sec:morphological_parameters}), and
more recent applications to X-ray maps can be found in
\citet{mantz:morphology,Nurgaliev_2017_X-Ray_morphology,
	Andrade-santos_2017_morphological_parameters_on_images} and \citet{Lovisari:x_ray_morphology}.

No morphological evaluation has been so far drawn from Sunyaev--Zel'dovich (SZ) maps observed in the millimetric band.
The SZ effect \citep{sz:70,sz:72} is produced by the Comptonization of the cosmic microwave background (CMB) photons 
from the interaction
with the energetic free electrons in the ICM and causes a redistribution of the energy of the CMB photons. This is
observed as a variation on the CMB background whose intensity depends on the observed frequency.
The CMB intensity variation is proportional to the integral of the electronic thermal
pressure of the ICM along the line of sight \citep[see e.g.][for a review]{carlstrom:sz}.
Therefore the SZ effect is linearly proportional to the electron number density. For this property, the SZ effect is a fundamental complementary tool to X-ray, since it probes more efficiently the outermost regions of galaxy clusters
\citep[see e.g.][]{roncarelli:icmstructure}.
Nowadays,  hundreds or even thousands of galaxy clusters observed through the SZ effect are available thanks to different surveys  carried out either  by ground-based  facilities , as in the case of the South Pole Telescope
\citep{Chang_2009_SPT_SZ_sky_survey,Staniszewski_et_al_2009_cluster_discovered_with_SZ,bleem:sptlatestcatalogue}
or the Atacama Cosmology Telescope \citep{act:presentation,Hasselfield_2013_ACT_three_seasons_of_data},
or space-based as in the case of the \textsl{Planck} satellite
\citep{planck:presentation, Planck_2014_SZ_catalogue, Planck_collab_2016:second_Planck_SZ_catalog}.
Clusters detected through the SZ effect do not show significant bias in terms of relaxation state. For instance,
\citet{Rossetti_2016_BCG_offset} shows that clusters detected by \textsl{Planck}
-- whose morphology is determined through the projected offset between the peak of the X-ray emission and the position
of the brightest cluster galaxy (BCG) -- are equally distributed between regular and disturbed objects.
On the contrary, in X-ray surveys the percentage of relaxed objects is $ \approx 74 $ per cent. This comparison suggests that
the observational selection effects can seriously influence the result. Hence, the importance of evaluating morphology from SZ catalogues that are
roughly mass-limited.

To infer cluster morphology from SZ maps one would require high sensitivity and high angular resolution. At present there are no SZ cluster catalogues
with such characteristics. However, the situation might change in the coming years.
Instruments like the currently operating MUSIC camera \citep{musiccamera:presentation} or MUSTANG-2
\citep{mustang2:presentation}, having maximum angular resolutions of $ \approx 30 $ and $ \approx 10 $ arcsec,
respectively, are examples of microwave detectors aimed at producing high-resolution cluster imaging through the SZ 
effect. Very promising results have also been recently reported with the 30-m telescope at the IRAM
observatory using the NIKA \citep{Monfardini_2010:NIKA_instrument}
and NIKA2 \citep{Calvo_2016_NIKA2_instrument,Adam_et_al_2017_NIKA_2} cameras, where the maps reach an
angular resolution of $ \approx 20 $ arcsec \citep{nika:firsttsz,Mayet_2017_NIKA2_observations,nika2:ruppin}.
Even better resolutions have been achieved at low frequencies with the use of interferometers, as shown in
\citet{alma:szat5arcsec}. Indeed, they report the SZ imaging of a galaxy cluster at 5 arcsec from observations with the
Atacama Large Millimeter Array \citep{alma:presentation}.
%\textcolor{red}{ %
Since SZ maps correspond to maps of the distribution of the thermal pressure, they are extremely valuable for the
investigation of cluster morphology
\citep[see][e.g.]{Wen_et_al_2013_Substructure_and_dynamical_state_of_2092_rich_clusters_of_galaxies_derived_from_photometric_data,
Cui_et_al_2017_MNRAS.464.2502C}.
For instance, in \citet{prokhorov:morphoSZsims} they highlight the effects produced by a violent merger on the SZ
imaging of a bullet-like simulated cluster, namely the presence of a cold substructure.
Morphology has some impact also on the scaling relation between the integrated SZ signal and the cluster mass,
as shown for simulated clusters \citep[e.g. in][]{Da_Silva_2001_impact_of_ICM_on_YM,
	McCarthy_2003_SZ_scaling_relation,Shaw_2008_SZ_scaling_relation}.
For instance, \citet{rumsey:SZscalingrelationvsmorpho} state that mergers induce small deviations from the canonical
self-similar predictions in SZ and X-ray scaling relations, in agreement with \citet{poole:merger_vs_scalingrel}.
%} %

Apart from these examples, there is not yet a detailed study on the morphology derived from observed or simulated 
SZ maps, or on the relation between SZ morphology and the cluster dynamical state.
The aims of this paper are therefore:
(1) to verify the feasibility of the application of some morphological parameters
typically used in X-ray on SZ maps,;
(2)  to determine their effectiveness in segregating the cluster
dynamical state (which in the case of simulated clusters is known a priori) and
(3) to evaluate their possible correlation with other relevant measures such as 
the hydrostatic mass bias or the X-ray morphological parameters.
For these goals, we use both existing and new parameters and we also combine them to derive a global parameter.

The paper is organized as follows.
In section \ref{sec:data set} we briefly describe the simulations and the data used in this analysis.
In section \ref{sec: 3D dynamical state segregation} we define the criteria used in the simulation to discriminate
the dynamical state, and provide a summary of the morphological parameters.
The efficiency and the stability of these parameters is tested in section \ref{sec: analysis}, together with a
comparison between their application to X-ray and SZ maps.
Finally, the correlation of the morphological indicators with the hydrostatic mass bias and with the projected shift
between the centre of mass (CM) and the centroid of the SZ map is investigated in sections \ref{sec: mas bias correlation}
and \ref{sec: projected shift correlation}.
We summarize our results and outline our conclusions in section~\ref{sec:conclusions}.
\section{Data set}
\label{sec:data set}
 %MODIFIED ER
 %ADDED CITATION RASIA+2008 TO XMAS2
 %FIX NUMBER OF SECTIONS
The analysis presented in this work is performed on simulated clusters taken from the
Marenostrum-MultiDark SImulations of galaxy Clusters (MUSIC\footnote{\url{http://music.ft.uam.es}})
\citep{Sembolini_et_al_2013:The_MUSIC_of_Galaxy_Clusters_I}.
The MUSIC project consists in two distinct sub-sets of re-simulated galaxy groups and clusters:
MUSIC-1, which is build on objects extracted from the MareNostrum simulation 
\citep{yepes:marenostrum}; MUSIC-2, that is build on systems selected within the MultiDark simulations
\citep{Prada_et_al_2012:Halo_concentrations_in_the_standard_Lambda_cold_dark_matter_cosmology}.
The resimulations of all clusters are based on the initial conditions generated by the 
zooming technique of \citet{Klypin_2001:zooming_technique_for_selecting_MUSIC_halos} and cover a spherical 
region centred on the redshift zero cluster with a radius of $ 6h^{-1} $Mpc. The Lagrangian regions are resimulated with the inclusion of the
baryonic physics and at higher resolution. 
The re-simulations are carried out using the TreePM+SPH \textsc{Gadget-2} code and include two different prescriptions
for the gas physics. The simplest follows the evolution of a non-radiative gas, while the other includes several physical
processes, such as cooling, UV photo-ionisation, stellar formation, and thermal and kinetic feedback processes
associated to supernovae explosions \citep[see details in][]{Sembolini_et_al_2013:The_MUSIC_of_Galaxy_Clusters_I}. 
We will refer to them as the NR and the CSF sub-sets, respectively. The final resolution is $ m_\textup{DM}= 9\times 10^{8} h^{-1} $ 
M$_\odot$ for the dark matter, and $ m_\textup{gas} = 1.9 \times 10^{8} h^{-1} $ M$_{\odot}$ for the
initial gas elements.

In this work we use a sample of 258 massive clusters, having virial mass
$M_{vir} > 5 \times 10^{14}h^{-1}$M$_{\odot}$ at $ z=0 $, extracted from the MUSIC-2 data-set.
We analyse clusters simulated with both ICM versions, and considered at four different times of their cosmic evolution,
namely at redshifts $ z = $ 0.43, 0.54, 0.67 and 0.82. The underlying cosmological model is that of the MultiDark parent 
simulation and it adopts the best-fit parameters from
\textsl{WMAP7}+BAO+SNI: $\Omega_{m}=0.27$, $\Omega_{b}=0.0469$, $\Omega_{\Lambda}=0.73$,
$\sigma_{8}=0.82$, $n=0.95$ and $h=0.7$
\citep[see][]{Komatsu_et_al_2011:Seven-year_Wilkinson_Microwave_Anisotropy_Probe_(WMAP)_Observations}.
This sample has been extensively analysed in the past, with focus on the baryon and SZ properties
\citep{Sembolini_et_al_2013:The_MUSIC_of_Galaxy_Clusters_I}, on the X-ray scaling relations
\citep{biffi:music2}, and on motions of both dark matter and gas \citep{Baldi_et_al_2017:MNRAS.465.2584B}.

%%%%%%%
\subsection{Sunyaev--Zel'dovich maps and X-ray data}
\label{sec:SZ and X maps}
The SZ effect can be separated into two components: the thermal component, produced by the random motion of the electrons in the ICM,
and the kinetic component, generated by the overall bulk motion of the cluster with respect to the CMB rest frame \citep[see e.g.][]{carlstrom:sz}. In this work we focus only on the former.
This produces a shift of the CMB brightness, $ I_\textup{CMB} $:
\begin{equation}
	\label{eqn:tsz}
	\frac{\Delta I}{I_\textup{CMB}} = \frac{x^4e^x}{(e^x-1)^2} \left[x \coth \left(\frac{x}{2} \right) - 4 \right] \ y
\end{equation}
where $ x = h_P \nu/(k_B T_\textup{CMB}) $, and $ \nu $ is the frequency of the radiation,
$ h_P $ is the Planck constant, $ k_B $ is the Boltzmann constant, and $ T_\textup{CMB} $ is the CMB temperature.
The $ y $ factor in equation~\eqref{eqn:tsz} is the Compton parameter, defined as:
\begin{equation}
	\label{eqn:comptonparameter}
	y =  \frac{\sigma_T k_B }{m_e c^2} \int n_e T_e d\ell
\end{equation}
where $ \sigma_T $ is the Thomson cross section, $ m_e c^2$ is the electron rest mass, 
$ n_e $ is the electron number density, $ T_e $ is the electron temperature, and the integration is performed over the
line of sight.
%} %
%\textcolor{red}{ %
In order to produce maps of the Compton parameter of our simulated clusters we refer to the discretized version of
the formula, proposed in \citet{Flores_Cacho_2009:generation_of_tSZE_maps_from_simulation}:
\begin{equation}
	\label{eqn:ydiscreto}
	y \simeq \frac{\sigma_T k_B}{m_e c^2} \sum_i \
	W_p\left( | \boldsymbol{r}_i - \boldsymbol{r}_{cm} |, h_s \right) n_{e,i} \ T_{e,i} \ \Delta \ell \ ,
\end{equation}
where the $ W_P $ function is the projected normalized spherical spline kernel of the simulation, i.e. the kernel
presented in \citet{monaghan:kernel}. The kernel depends on the SPH smoothing length $ h_s $ and is evaluated
at the radial distance of the $ i $-th particle with respect to the  cluster centre of mass, $ | \boldsymbol{r}_i - 
\boldsymbol{r}_{cm} | $.
The sum in equation~\eqref{eqn:ydiscreto} extends to all gas particles located along the line of sight up to a maximum
distance of $ 1.5R_{vir} $ from the cluster centre, being $ R_{vir} $ the virial radius.
%%
%\begin{equation}
%		W (r, h_s) = \frac{8}{\pi h_s^3} \ 
%			\resizebox{0.65\columnwidth}{!}{$ %
%			\begin{cases}
%				1 - 6(r/h_s)^2 + 6(r/h_s)^3, &  0 \leq (r/h_s) \leq 0.5 \\
%				2\left(1 - (r/h_s)\right)^3, & 0.5 \leq (r/h_s) \leq 1 \\
%				0, & (r/h_s) > 1
%			\end{cases}
%		$}
%\end{equation}
%%
%where $ h_s $ is the SPH smoothing length.
The side of the maps has a physical size of 10 Mpc, corresponding to $ \sim $ 3.4 times the mean virial radius of
the sample.
This extension is comparable with the maximum radius that can be probed with SZ measurements of large
clusters, which e.g. for \textsl{Planck} is of the order of a few virial radii
%For the purpose of studying the SZ signal from intermediate to high redshifts (as the ones selected here, from
%0.43 to 0.82), such extension of the maps is significantly larger than the maximum physical radius which can be
%explored through SZ measurements, which for large clusters is of the order of the virial radius and beyond
\citep{planck:pressure_rvir}.
Each pixel is equivalent to 10 kpc.
In angular distances, given the cosmological parameters adopted in the simulation, the field of view and the pixel 
resolution correspond to (29.6, 26.1, 23.6, 21.8) arcmin and (1.8, 1.6, 1.4, 1.3) arcsec at the four redshifts ($z=0.43, 
0.54, 0.67, 0.82$), respectively.
Such large fields of view could be covered with large mosaics of detectors and multiple observational runs.
For example, one would need 4 to 9 pointings with the
MUSIC camera, which has a field of view of $ \approx 10 $ arcmin.
The angular resolution of our maps, instead, is not achievable by any current instrument, that at best reaches
5 arcsec for interferometric measurements (e.g. with ALMA), and about 20 arcsec for single-dish measurements
(e.g. with the NIKA2 camera).
%} %
In section~\ref{subsec:angularresolutions}, we will degenerate the SZ signal to reproduce the
angular resolution of three examples of existing telescopes for microwave astronomy, having diameters of
1.5, 10 and 30 m, respectively.
In this work we do not consider specific observational features such the instrumental noise
or any contamination with astrophysical origins. These will be properly taken into account in a forthcoming work which will address the capabilities of
a specific experiment.
The noise, indeed, typically shows a significant pixel-to-pixel correlation, and it is an intrinsic characteristic of
the instrument. Astrophysical sources of contamination also depend on the instrument, in particular on the
observed frequencies.
In this study we refer to noiseless and maps without any contamination to be as general as possible.
%} %

In section~\ref{subsubsec:SZvsXrays} we compare the morphological parameters derived from
SZ with those obtained from X-ray data.
We use two sub-samples of the non-radiative clusters that have been selected in
\citet{Meneghetti_et_al_2014:The_MUSIC_of_Clash:Predictions_on_the_Concentration-Mass_Relation} to
morphologically match the CLASH sample observed by \textsl{Chandra} and \textsl{XMM-Newton} in X-ray
\citep{CLASH:presentation}. These sub-samples are constituted by 79 clusters at $ z=0.43 $ and 86 clusters
at $ z=0.67 $. Their X-ray maps were produced using the
X-MAS software package \citep[see][]{Gardini_2004_X_MAS,rasia:xmas2008} to mimic ACIS-S3 \textsl{Chandra}
observations, with a field of view of 8.3 arcmin and angular resolution of 0.5 arcsec.
%} %
\section{Determination of the dynamical state}
%MODIFIED BY ER
%THE BEGINNING IS CHANGE (FORM NOT CONTENT)
\label{sec: 3D dynamical state segregation}
Before presenting the morphological parameters, we introduce the indicators of the cluster dynamical state 
that, in simulations, can be measured in a quantitative way from several estimators.
%To test the classification obtained from the application of the morphological parameters, the dynamical
%state of the clusters in the sample should also be evaluated.
%With respect to observations, simulations provide the significant advantage of having the possibility to extract this
%information from the data. Indeed, the relaxation state of a cluster can be quantified from several estimators.
%\textcolor{red}{ %
For instance, one of the most used indicators in the literature is the ratio between the
kinetic energy, $T$, and the potential energy, $W$, of the system measured within the virial radius. The ratio
is expressed as $(2T-E_S)/|W|$, where 
$ E_S $ is the surface pressure energy evaluated at the same virial radius.
The cluster is considered relaxed when the ratio is lower than 1.35 \citep[see e.g.][]{Neto_2007_threshold_virial_ratio, 
Ludlow_et_al_2012_dynamical_state}.
Even if broadly adopted, in this paper we prefer to avoid its usage since several works showed that it is often unreliable
\citep[see e.g.][]{Sembolini_et_al_2014:The_MUSIC_of_Galaxy_Clusters-III, Klypin_et_al_2016:MNRAS.457.4340K}.
Another criterion is based on the ratio of the gas velocity dispersion, $\sigma$, over the theoretical velocity dispersion, 
$\sigma_t$  \citep{Cui_et_al_2017_MNRAS.464.2502C}.
This indicator, often expressed as $\zeta=\sigma/\sigma_t$, could be applied to optical data, but it is still not clear whether the threshold that discriminates between relaxed and disturbed objects is mass dependent. 
 
%We evaluate the dynamical state of our simulated sample via the two parameters described in the following.
% For this reason, we prefer to use the two indicators described below.
%} %

\subsection{Indicators of the dynamical state}
\label{subsec:3Dindicators}
In this work, the cluster dynamical state is quantified through the following two parameters, derived from the 3D information 
of the simulated cluster. For this reason, we also called them 3D indicators:
\begin{itemize}
	\item the ratio between the mass of the biggest sub-structure and the cluster mass evaluated within the virial radius 
	$M_{sub}/M_{vir}$. 
	Some variants of this parameter exist, including the ratio between the mass of all substructures and the total cluster mass
	\citep[see e.g.][]{Biffi_et_al_2016:On_the_Nature_of_Hydrostatic_Equilibrium_in_Galaxy_Clusters,
		Meneghetti_et_al_2014:The_MUSIC_of_Clash:Predictions_on_the_Concentration-Mass_Relation}, however,
	 we refer to its simplest definition, as e.g. in \citet{Sembolini_et_al_2014:The_MUSIC_of_Galaxy_Clusters-III};
	\item the offset $\Delta_{r}$ between the position of the peak of the density distribution, $\mathbf{r}_{\delta}$, and 
	the position of the centre of mass of the cluster, $\mathbf{r}_{cm}$, normalized to the virial radius $R_{vir}$: 
	\begin{equation}
	\Delta_{r}=\frac{|\mathbf{r}_{\delta}-\mathbf{r}_{cm}|}{R_{vir}} \ .
	\label{eq:center of mass offset}
	\end{equation}
\end{itemize}
We classify the cluster as relaxed if both indicators are simultaneously smaller than a certain threshold, that we fix equal to 0.1 in both cases.
In literature, different thresholds are adopted for DM only simulations
\citep[see e.g.][]{Maccio_et_al_2007:Concentration_spin_and_shape_of_dark_matter_haloes,
	D'Onghia_et_al_2007:Do_mergers_spin-up_dark_matter_haloes?}.
%we choose to adopt 0.10 for both $\Delta_r$ and $M_{sub}/M_{vir}$, 
In particular, for the offset parameter, it is often used a smaller value for the thresold. However, we prefer to increase it to 0.1 to accomodate the 
effect of baryons that reduce the displacement due their collisional nature.
%by the fact that the introduction of baryons is know to enlarge the offset.  
%which is a 
%reasonable value for hydrodynamic simulations, since this type of simulation is known to give higher values of
%$\Delta_{r}$ with respect to dark matter-only simulations.
%In particular, we impose that a cluster is dynamically relaxed if both parameters are below the threshold value of
%0.10 at the same time.
%We show the resulting percentages of 
The fraction of relaxed and disturbed clusters is shown in Table~\ref{tab:3D classification 
percentages}, for all the analysed data sets. No significant dependence on redshift or ICM physics
is seen in our data. Our sample has about 55 percent of relaxed objects.
\begin{table}
	\centering
	\caption{Percentages of clusters classified as relaxed or disturbed for the four considered redshifts and
		the two ICM physics.}
	\label{tab:3D classification percentages}
	\begin{tabular}{lcccc} % three columns, alignment for each
		\hline
		$z$ & \multicolumn{2}{c}{CSF} & \multicolumn{2}{c}{NR} \\
		& relaxed & disturbed & relaxed & disturbed \\
		\hline
		0.43 & 56\% & 44\% & 55\% & 45\% \\
		0.54 & 53\% & 47\% & 53\% & 47\% \\
		0.67 & 56\% & 44\% & 55\% & 45\% \\
		0.82 & 54\% & 46\% & 53\% & 47\% \\
		\hline
	\end{tabular}
\end{table}

%%%%%%%%%%%
\subsection{Morphological parameters}
\label{sec:morphological_parameters}
In the following we describe the morphological parameters analysed in this work:
the \emph{asymmetry parameter}, the \emph{fluctuation parameter}, the \emph{light concentration parameter},
the \emph{third-order power ratio parameter}, the \emph{centroid shift parameter}, the \emph{strip parameter},
the \emph{Gaussian fit parameter} and a \emph{combined parameter}.
The first five indicators are taken from X-ray morphological studies, while we 
introduce here the remaining parameters.
Here, we will discuss their expected behaviour in 
the case of relaxed or disturbed clusters. 

All the parameters refer to the centroid of the 
analysed SZ map as center and we computed them inside different values of the
\emph{aperture radius}, $R_{ap}$, equivalent to  0.25, 0.50, 0.75 and 1.00 times $R_{vir}$.
We will discuss which aperture works best for each parameter
(see section \ref{sec: analysis}).

%%%%%%%%%
\subsection*{Asymmetry parameter, $A_{\theta}$}
This parameter, originally introduced by \citet{Schade1995:asymmetryparameter}, is based on the
normalized difference between the original SZ map, $I$, and the rotated map, $R_{\theta}$, being $\theta$
the rotation angle:
\begin{equation}
A_{\theta}=\frac{\sum\limits_{r<R_{ap}} | I - R_{\theta} |}{\sum\limits_{r<R_{ap}} I} \ ,
\label{eq:asymmetry_param_definition}
\end{equation}
where the sums are extended to all pixels within $R_{ap}$.
We computed four different versions of this parameter,  $A_{x}$, $A_{y}$, $A_{\pi}$, and $A_{\pi/2}$. 
For $A_{x}$ and $A_{y}$ we consider as  $R_{\theta}$ the flipped image along the $ x $ or the $ y $ axes, 
respectively \citep[as in][]{Rasia_et_al_2013:X_ray_Morphological_Estimators_for_Galaxy_Clusters}.
Low values of $A_{\theta}$ indicates relaxed clusters.
With the purpose of making a final classification, we use the rotation angle corresponding
to the maximum value of $A_{\theta}$.

%%%%%%%%
\subsection*{Fluctuation parameter, $F$}
The \emph{fluctuation parameter}, introduced in \citet{Conselice2003:fluctuationparameter}, is defined 
similarly to the asymmetry parameter. Namely, it is expressed as the normalized difference between an original image, 
$I$, and its  Gaussian-smoothed version, $B$:
\begin{equation}
F=\frac{\sum\limits_{r<R_{ap}} | I - B |}{\sum\limits_{r<R_{ap}} I} \ .
\label{eq:fluctuation_param_definition}
\end{equation}
Various versions of this parameter exist: negative residuals are ignored in
\citet{Conselice2003:fluctuationparameter}, or the absolute value are not considered in 
\citet{Okabe2010:fluctuationparameter}, or different values for the full-width-half-maximum (FWHM) Gaussian filter are used.
We keep the absolute values to account both negative and positive residuals. Furthermore, we consider
10 equally-spaced FWHM values, from 0.05$R_{vir}$ to 0.5$R_{vir}$ to evaluate the most effective choice.
Regular clusters are expected to present low values for the fluctuation parameter.

%%%%%%%%%%%%%%
\subsection*{Light concentration  parameter, $c$}
This parameter was introduced by \citet{Santos_et_al_2008:Searching_for_cool_core_clusters_at_high_redshift} 
with the purpose of segregating cool core and non-cool core clusters
\citep[see also][]{Cassano2010:morphologicalparameters,
Rasia_et_al_2013:X_ray_Morphological_Estimators_for_Galaxy_Clusters}. 
These analyses, based on X-ray maps, reach at best $R_{500}$\footnote{$R_{500}$ is the radius
of a spherical volume enclosing a density 500 times larger than the critical density.} (and only in few cases they
go beyond). Here we use a more general mathematical
formula given by the ratio of the surface brightness computed within a radius $r_2$ and 
the one evaluated within a more central region of radius $r_1 < r_2$:
\begin{equation}
c=\frac{\int_{0}^{r_1}S(r)dr}{\int_{0}^{r_2}S(r)dr} \ .
\label{eq:light_concentration_param_definition}
\end{equation}
We choose the inner and outer radii as fractions of the virial radius, in order to avoid any dependence on redshift 
when assuming a fixed physical aperture 
\citep[as discussed in][]{Hallman2011:why-not-a-physical-radius-in-light-concentration-parameter} as it was done in \citet{Santos_et_al_2008:Searching_for_cool_core_clusters_at_high_redshift} .
In particular, we set $r_2 = R_{ap}$, and we use ten different values of $r_1$, uniformly sampled
between 0.1 and 1.0 times $ r_2 $.
%The final values corresponding to the most efficient configuration, are chosen as described in section \ref{sec: analysis}}.
We expect higher values of $c$ for relaxed clusters, since in this case the surface brightness is peaked near
the cluster centre, and lower values for disturbed ones, due to their irregular shape and the possible presence of
structures located far from the centre.

%%%%%%%%%%%%%%
\subsection*{Third-order power ratio parameter, $P_3/P_0$}
The first definition of the m-$th$ order power ratio, $P_m/P_0$, was given by 
\citet{Buote_et_al_1995:Quantifying_the_Morphologies_and_Dynamical_Evolution_of_Galaxy_Clusters_I_The_Method}:
\begin{equation}
\frac{P_m}{P_0}=\frac{a_{m}^{2}+b_{m}^{2}}{2m^{2}R^{2}_{ap}a_{0}\ln(R_{ap})} \ ,
\label{eq:power_ratio_param_definition_1}
\end{equation}
where the coefficients $a_m$ and $b_m$ are, respectively, defined as:
\begin{equation}
a_{m}(R_{ap})=\int\limits_{r\leq R_{ap}}S(r, \phi) r^m \cos(m\phi) dr d\phi \ ,
\label{eq:power_ratio_param_definition_a_m}
\end{equation}
\begin{equation}
b_{m}(R_{ap})=\int\limits_{r\leq R_{ap}}S(r, \phi) r^m \sin(m\phi) dr d\phi \ ,
\label{eq:power_ratio_param_definition_a_b}
\end{equation}
and $S(r,\phi)$ is the surface brightness expressed as a function of the projected radius and azimuthal 
angle, $ \phi $.
Following this definition, we compute the third-order power ratio $P_3/P_0$ considering $m=3$, and we take its decimal logarithm.
This parameter is one among the most efficient in X-ray
\citep[see e.g.][]{Rasia_et_al_2013:X_ray_Morphological_Estimators_for_Galaxy_Clusters,
Lovisari:x_ray_morphology}.
We measure $P_3/P_0$ using four different aperture radii, expressed as fractions of the virial 
radius, and then consider its maximum value
to better identify clusters with sub-structures or irregular shape which are associated to high values of the power ratio.

%%%%%%%%%%%%%%%%%
\subsection*{Centroid shift parameter, $w$}
The centroid shift parameter is a measure of how much the 
%purely geometrical, since it is derived from the computation of the
centroid of the map with different circular sub-apertures changes.
 Once the centroid within $R_{ap}$ is computed, a new aperture radius is defined and the respective centroid is found. The operation is repeated for $N$ sub-apertures. The centroid shift is
then defined as the normalized standard deviation of all $ \Delta_i $ separations:
\begin{equation}
w=\frac{1}{R_{ap}}\sqrt{\frac{\sum(\Delta_i - <\Delta_i>)^{2}}{N-1}}
\label{eq:centroid_shift}
\end{equation}
where $<\Delta_i>$ is their mean value.
This parameter has been largely applied on X-ray maps in several works, altough some variations in its definition were considered. 
 Generally it was found to be an efficient parameter to discriminate the cluster dynamical state
\citep[see e.g.][]{Mohr_et_at_1993:An_X-ray_method_for_detecting_substructure_in_galaxy_clusters_[centroid_shift],
O'Hara_et_al_2006:centroid_shift,Poole_et_at_2006:The_impact_of_mergers_on_relaxed_X-ray_clusters_[centroid_shift],
Maughan_et_al_2008:Images_Structural_Properties_and_Metal_Abundances_of_Galaxy_Clusters_Observed_with_Chandra_[centroid_shift],
Ventimiglia_et_al_2008:Substructure_and_Scatter_in_the_Mass-Temperature_Relations_of_Simulated_Clusters_[centroid_shift],
Jeltema_et_al_2008:Cluster_Structure_in_Cosmological_Simulations...[centroid_shift],
Boehringer_et_al_2010:Substructure_of_the_galaxy_clusters_in_the_REXCESS_sample_[centroid_shift],
Weissmann_et_al_2013:Studying_the_properties_of_galaxy_cluster_morphology_estimators_[centroid_shift]}.
The presence of a substructure will increase the value of the centroid shift, therefore we expect low values for relaxed
clusters. Even though, this might not be the case in the presence of symmetric substructures that would be unidentified by $w$.

%%%%%%%%%%%%%%%%%
\subsection*{Strip parameter, $S$}
%The first new parameter that we define is the strip parameter.
We call ``strip'' a profile extracted from the SZ map and
passing through its centroid.
The strip parameter is the sum of the pixel-to-pixel difference between
couples of different strips, $S_{i}$ and $S_{j}$, at $ N $ total different angles in absolute value. In order to obtain a value of $S$
between 0 and 1, this sum is 
normalised by the maximum strip integral and by the number of the strip pairs considered:
\begin{equation}
S=\frac{\sum_{\substack{i,j\\j<i}}|S_i(r) - S_j(r)|}{\frac{N(N-1)}{2}\max[\int\limits_{R_{ap}} S_i(r) dr]} \ .
\label{eq:strip_parameter_analytical_definition}
\end{equation}
As usual, $ R_{ap} $ indicates the aperture radius, i.e. the maximum radius within which the
integration of the strips is performed.
The main advantage of using this parameter is the possibility to use many non-repeated combinations of strips.
Indeed, in this way the parameter quantifies the different contributions to the overall symmetry of the
cluster SZ map coming from multiple angles. %, taking only a 45-degrees rotation of the map.?????????
In our case, performing a single rotation of the SZ maps, we take four strips selected with angles equal to 0$ ^{\circ} $, 45$ ^{\circ} $, 90$ ^{\circ} $ and 135$ ^{\circ} $, for a total of six computed differences.
%\textcolor{red}{ %
We have tested the effects of using $ N > 4 $ angles, finding that the dynamic range of this parameter is slightly reduced with increasing number of strips. Nevertheless, the overlap between the populations of relaxed and disturbed objects remains substantially unchanged, allowing us to choose a small number of strips to reduce the computational time without significantly affecting the final results. 
%We verified that a different number of strips does not substantially improve the results.
%} %
Disturbed clusters are expected to show higher values of the parameter because of the presence
of sub-structures (visible as off-centre peaks in the strips) or possible asymmetries. 
%The opposite situations occurs
%for relaxed clusters, whose strips should appear mainly superimposed.

%%%%%% UNTIL HERE %%%%%%%%%%%
\subsection*{Gaussian fit parameter, $G$}

The Gaussian fit parameter, $G$, is based on a two-dimensional Gaussian fitting of the SZ maps applied within the 
aperture radius.
%The second parameter we introduce here, the Gaussian fit parameter  is purely based on morphology.
%\textcolor{red}{
%To compute this parameter, we fit the SZ maps within a radius $R_{ap}$ to a two-dimensional
%Gaussian model, that 
The Gaussian model used can be written in terms of the $ x $ and $ y $ coordinates as:
\begin{equation}
	\begin{split}
		& f(x,y) =  z_0 + \\
		& A \exp \left\lbrace - \left[ a \left( x - x_0 \right)^2 + 2b \left( x - x_0 \right) \left( y - y_0 \right)^2 +
		 c \left( y - y_0 \right) \right] \right\rbrace \ ,
	\end{split}
	\label{eq:gaussian_fit_parameter_fitting_function}
\end{equation}
where $a$, $b$ and $c$ are constants defined as:
\begin{equation}
\begin{split}
& a = \frac{\cos^2\theta}{2\sigma_x^2} + \frac{\sin^2\theta}{2\sigma_y^2} \\
& b = -\frac{\sin 2\theta}{4\sigma_x^2} + \frac{\sin 2\theta}{4\sigma_y^2} \\
& c = \frac{\sin^2\theta}{2\sigma_x^2} + \frac{\cos^2\theta}{2\sigma_y^2} \ .
\end{split}
\label{eq:gaussian_fit_parameter_a_b_c_parameters}
\end{equation}
The best-fitting parameters obtained from the procedure are the following: 
the angle $\theta$ between the axes of the map and those of the bi-dimensional Gaussian
distribution;
%map coordinate axes and the ones referred to the
%best-fit Gaussian map, 
the coordinates of the peak of the model map, $x_0$ and $y_0$; its amplitude $A$; the 
offset $z_0$ and the two standard deviations, $\sigma_x$ and $\sigma_y$.
%Given the smallest and the largest among the aforementioned standard deviations taken from the
%best fit ($\sigma_{min}$ and $\sigma_{max}$, respectively), 
The $G$ parameter is just defined as the ratio:
%}
%
\begin{equation}
G=\frac{\sigma_{min}}{\sigma_{max}},
\label{eq:gaussian_fit_parameter}
\end{equation}
where $\sigma_{min}$ ($\sigma_{max}$) denotes the smallest (largest) value between $\sigma_x$ and $\sigma_y$
We expect lower values of $G$ for disturbed clusters, which should show asymmetric shapes,
resulting in significantly different standard deviations along the $ x $ and $ y $ direction in the
Gaussian fit. Instead, $G$ should be close to $1$ for regular clusters. It should be stressed that, similarly to the 
case of the centroid shift, this parameter could lead to a misclassification of disturbed clusters in presence of symmetrically
distributed sub-structures. It should also be noted that the aperture radius within which the fit is computed should
be sufficiently large, in order to take into account the presence of possible sub-structures located far from the 
centre.
\begin{figure*}
	\centering
	\subfloat
	{\includegraphics[width=1\columnwidth]{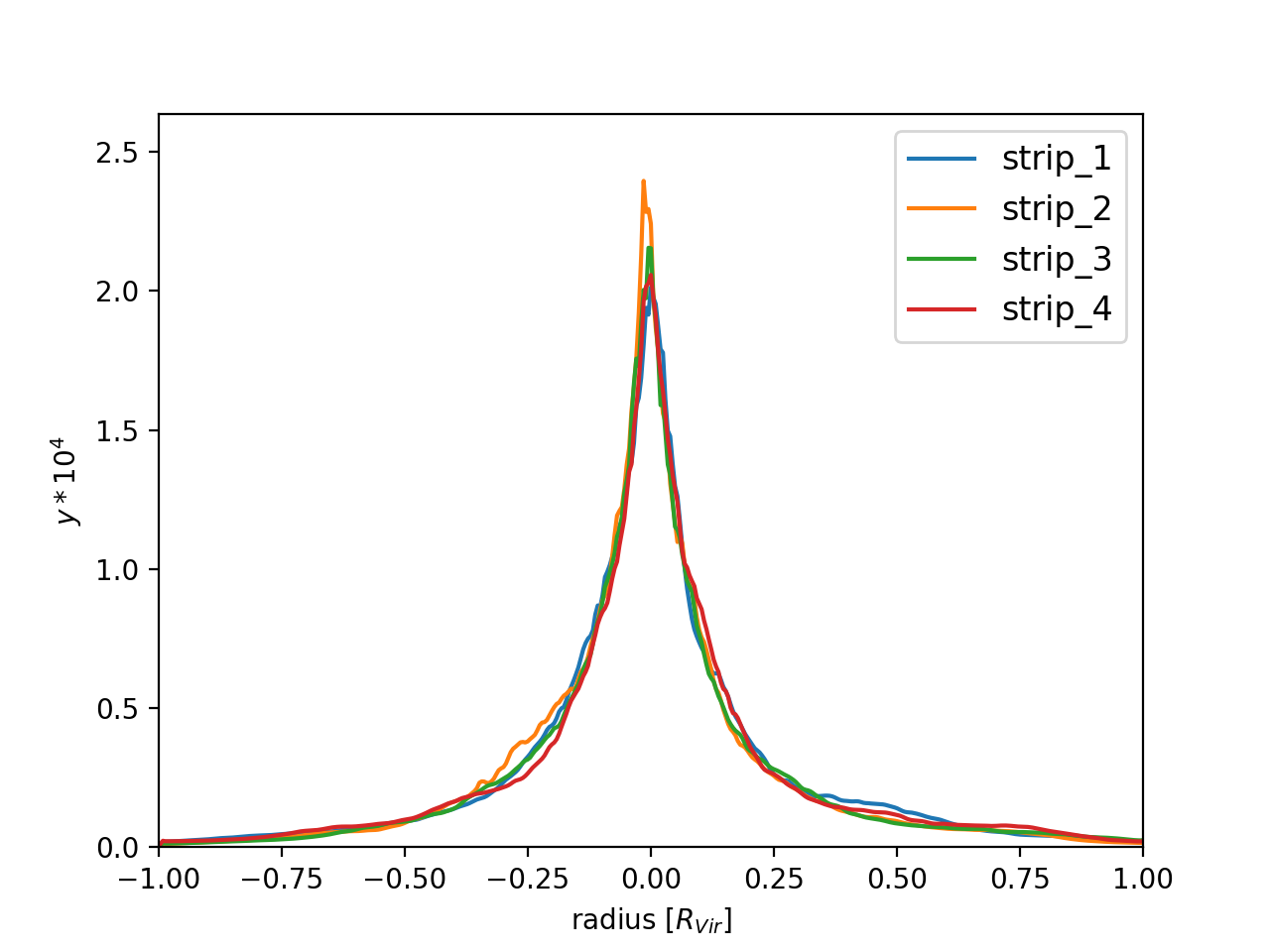}}
	\subfloat
	{\includegraphics[width=1\columnwidth]{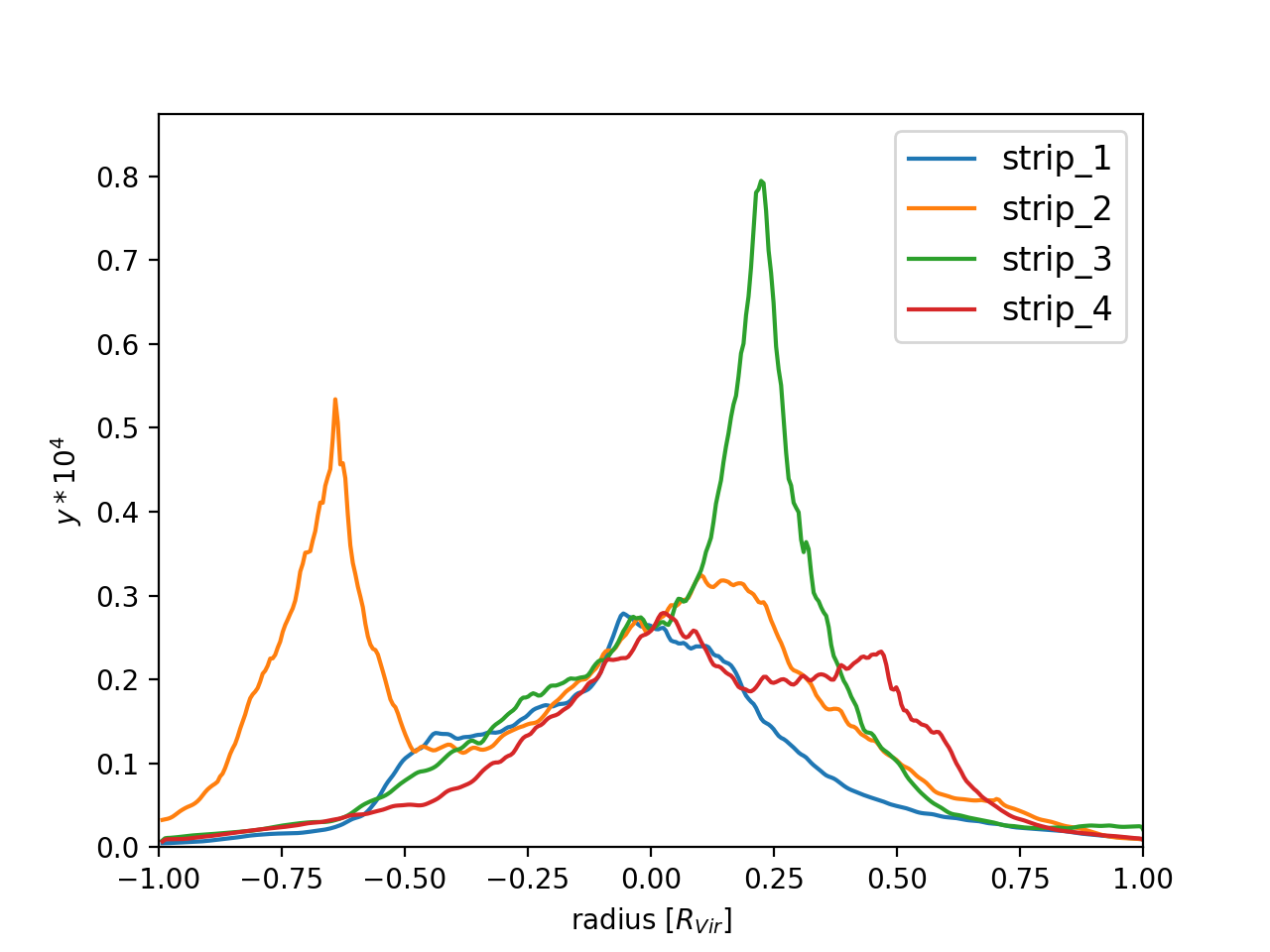}}
	\caption{Strips passing through the centroid of the SZ maps of a relaxed cluster (\#7, left panel) and of an unrelaxed 
		one (\#27, right panel) for the four chosen orientations (see text for the description) at redshift $z=0.54$ for the
		CSF flavour. The corresponding values of the $ S $ parameter for the two clusters are 0.09 and 0.40,
		respectively.}
	\label{fig: S parameter brightness profiles}   
\end{figure*}
\begin{figure*}
	\centering
	\subfloat
	{\includegraphics[width=1\columnwidth]{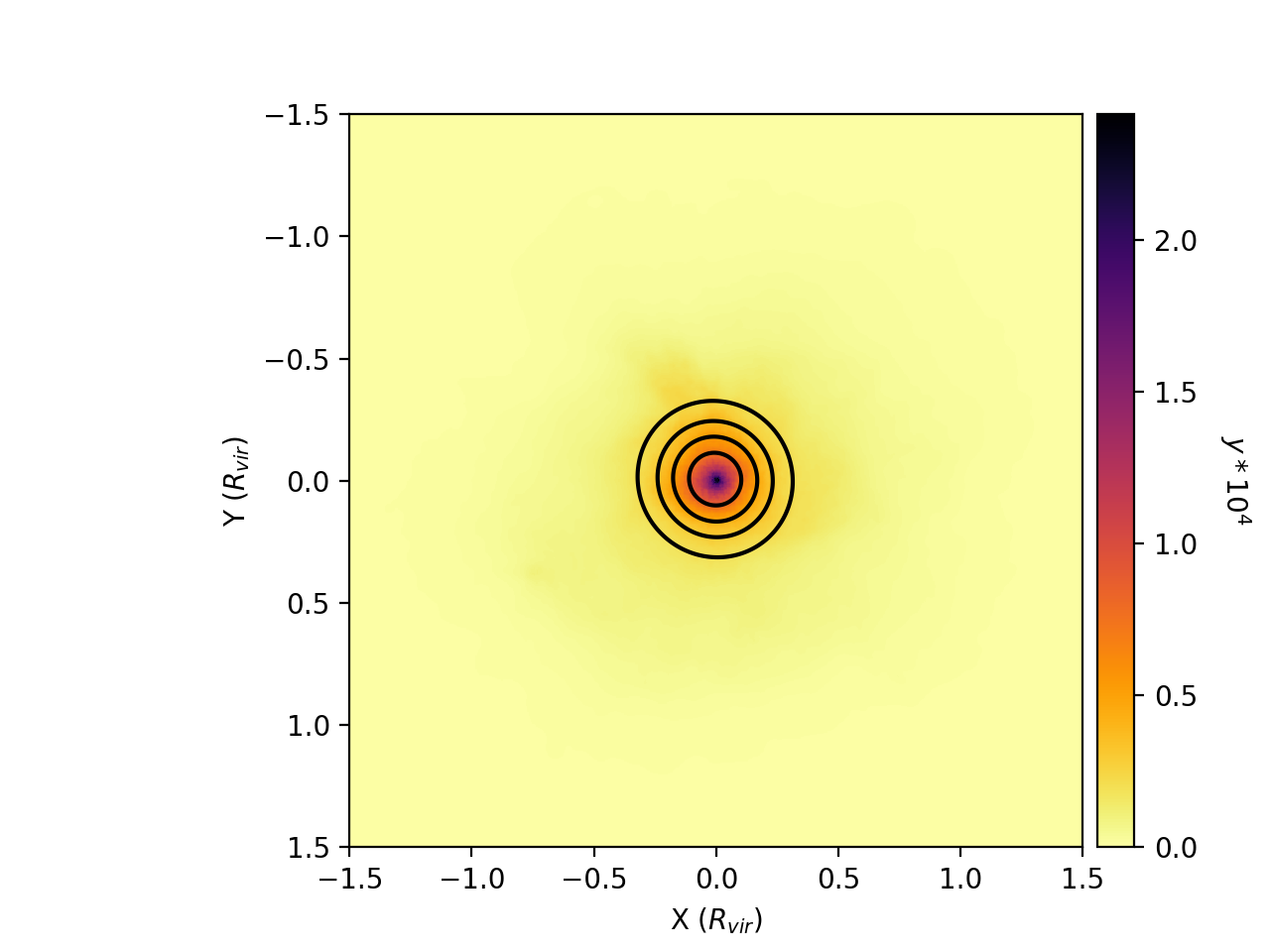}}
	\hspace{5mm}
	\subfloat
	{\includegraphics[width=1\columnwidth]{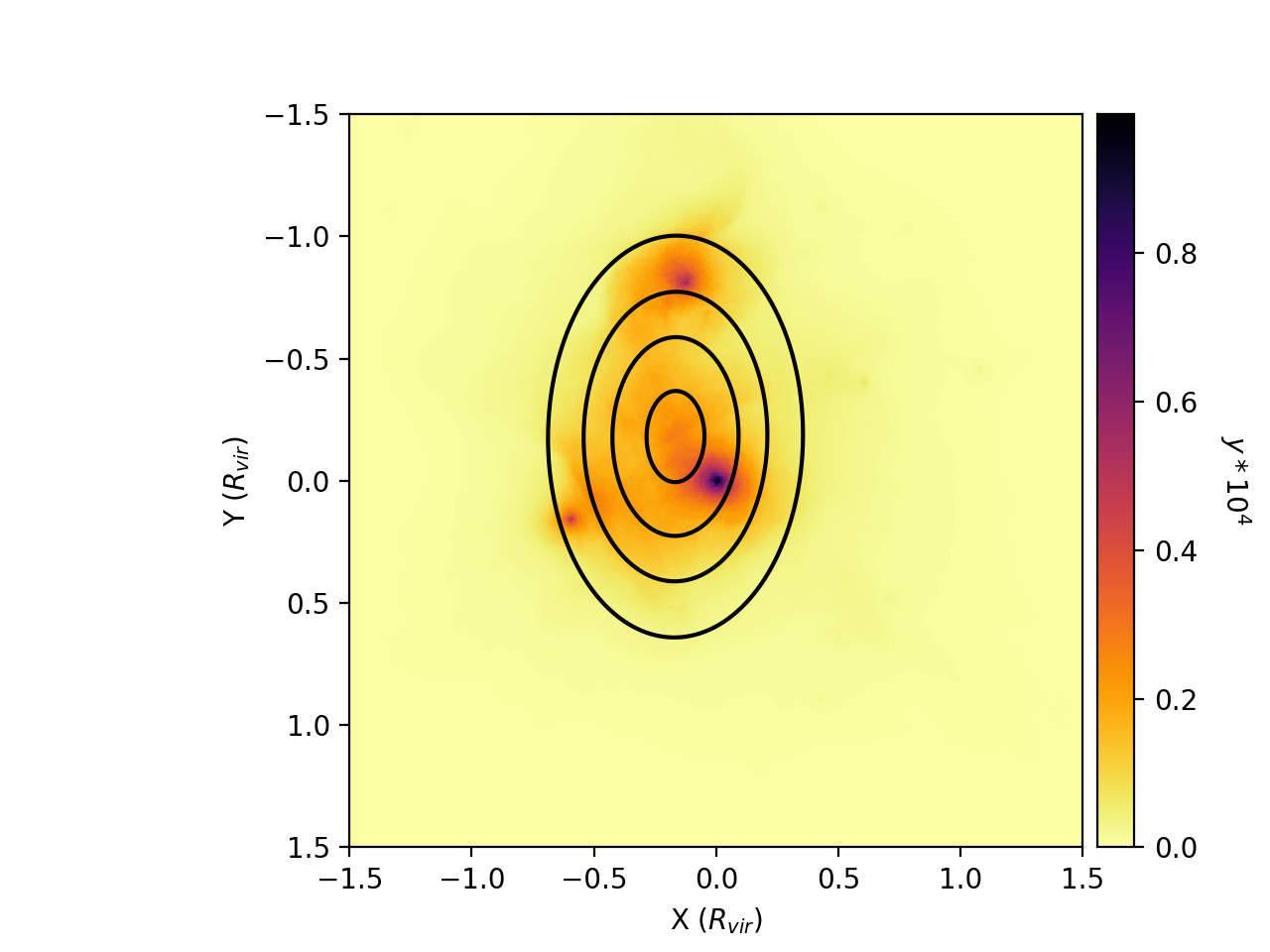}}
	\caption{Best Gaussian fit contour lines (at 20, 40, 60 and 80 per cent of the maximum $y$ value), resulting
		from the computation of the G parameter. Contours are superimposed to the SZ maps of the example relaxed
		(\#7, left panel) and disturbed cluster (\#27, right panel) at redshift $z=0.54$ from the CSF sub-set.
		The corresponding values of the $ G $ parameter are 0.97 and 0.63 for the two clusters, respectively.}
	\label{fig: G parameter gaussian fits examples}   
\end{figure*}
%

%\textcolor{red}{%
\vspace{0.2cm}
To illustrate how the new strip and Gaussian parameters work, we show their different behaviours
for an example of relaxed and disturbed cluster (\#7 and \#27, respectively) from
the radiative sub-set at $ z=0.54 $.
Fig.~\ref{fig: S parameter brightness profiles} shows the strips defined as
above. As expected, the four profiles for the relaxed cluster are similar at all radii ($S=0.09$).
On the contrary multiple off-centre peaks characterise the profiles of the disturbed cluster 
in correspondence of of sub-structures ($S=0.4$).
Fig.~\ref{fig: G parameter gaussian fits examples} illustrates the maps of the two clusters and the contour lines
representing the Gaussian fit to the maps. The difference between the relaxed and the unrelaxed cluster can be seen
from the shapes of the contour lines, which are nearly circular in the first case ($G=0.97$) and elliptical in the second case ($G=0.63$).
%}

%%%%%%%%%%%
\subsection*{Combined parameter, $M$ }
\label{sec: the M parameter}
%With the aim of taking into account all the morphological parameters into a global one, and to
%maximize the discrimination efficiency, 
Finally, we introduce a combined parameter called $M$ 
\citep[see also][]{Meneghetti_et_al_2014:The_MUSIC_of_Clash:Predictions_on_the_Concentration-Mass_Relation,
Rasia_et_al_2013:X_ray_Morphological_Estimators_for_Galaxy_Clusters}, where each of the
previously defined parameters, denoted generically as $ V_i $, contributes to $ M $ according to a weight, $ 
W_i $, related to the efficiency of $V_i$ itself in discriminating the dynamical state, as will be detailed later in
section~\ref{subsec: results M parameter}.
The analytical definition of the combined parameter is thus:
\begin{equation}
	M = \frac{1}{ \sum_{i} W_i} \left( \sum_{i} W_i \frac{\log_{10}(V_i^{\alpha_i}) - 
	<\log_{10}(V_i^{\alpha_i})>}{\sigma_{\log_{10}(V_i^{\alpha_i})}} \right)
	\label{eq:M_parameter}
\end{equation}
where the sums extend over all the parameters, and $\alpha_i$ is equal to +1 when disturbed clusters are associated to
large values of $ V_i $ (e.g. as in the case of $A$), otherwise it is equal to -1 (e.g. as in the case of $c$).
The  brackets, $< >$, indicate the average computed over all the clusters, and $ \sigma $ is the standard 
deviation. By definition, we expect negative values for the relaxed clusters,
and positive values for the disturbed ones.
%\begin{equation}
%\begin{split}
%M = A_1 \frac{log_{10}(A) - <log_{10}(A)>}{\sigma_{log_{10}(A)}} + A_2 \frac{log_{10}(F) - 
%<log_{10}(F)>}{\sigma_{log_{10}(F)}} \\ 
%+ A_3 \frac{log_{10}(1/c) - <log_{10}(1/c)>}{\sigma_{log_{10}(1/c)}} + A_4 \frac{log_{10}(P) - 
%<log_{10}(P)>}{\sigma_{log_{10}(P)}} \\ 
%+ A_5 \frac{log_{10}(w) - <log_{10}(w)>}{\sigma_{log_{10}(w)}} + A_6 \frac{log_{10}(1/G) - 
%<log_{10}(1/G)>}{\sigma_{log_{10}(1/G)}} \\ 
%+ A_7 \frac{log_{10}(S) - <log_{10}(S)>}{\sigma_{log_{10}(S)}}
%\end{split}
%\label{eq:M_parameter_extended}
%\end{equation}
\section{Results}
%MODIFIED BY E.R.
% NEEDS TO BE READ AGAIN
%
\label{sec: analysis}
In this section we first discuss the efficiency of the parameters described in 
section \ref{sec:morphological_parameters}, then
we analyse their stability by varying the observer line of sight, and thus considering multiple projections of the same cluster,
and by changing the angular resolutions of the maps.
%}

%%%%%%%%%%
\subsection{Application of the single parameters}
\label{sec: x ray comparison}
We quantify the efficiency of the morphological parameters with 
a Kolmogorov-Smirnov test (KS test hereafter) on the distributions of the two populations of relaxed and 
disturbed objects, as identified from the 3D indicators of the cluster dynamical state (see section~\ref{subsec:3Dindicators}). 
From this test, we obtain the probability, $ p $, that they belong to the same sample. Low values of this probability indicate an 
efficient discrimination between the two populations. 

For all the parameters, the efficiency depends on the aperture radius $R_{ap}$ and in few cases also on the inner radius, such as for the light concentration ratio, or on the FWHM for the fluctuation parameter.
As described in the previous section we consider multiple values for all these quantities, and we finally chose those
that correspond to the lower value of the probability $ p $, averaged over the four
redshifts and two sub-samples. The fluctuation parameter returned contradictory results among the redshifts and physics 
demonstrating that it is not a stable parameter. For this reason, it will be discarded in the rest of the analysis.
The minimum probability values for each parameter are listed in Table~\ref{tab:superimposition_percentages_allparams}, where we also present the superimposition percentage, $s_p$, of the distributions of the two populations and the most efficient aperture
radius.
The latter turns to be $ R_{ap} = 0.75 R_{vir} $ for the asymmetry 
parameter, $ R_{ap} = 0.25 R_{vir} $ for the light concentration ratio, and $ R_{ap} = R_{vir} $ for all the others.

In general, we find small values for the KS probability, so we can conclude that the relaxed and disturbed populations do not coincide. This implies 
a good agreement between the dynamical state expected from the 3D indicators and the one
inferred from the parameters. 
Since the results are similar for the NR and the CSF flavours, we conclude that the
gas radiative processes do not have significant impact on the SZ maps on which the parameters are computed
\citep[consistently with e.g.][]{Motl:szsuperior}.
Differences are also negligible, within few per cent, in terms of redshift variation.

The light concentration ratio, for which the best inner radius is $ r_1 = 0.05R_{vir} $,
is the one showing less overlap between the two classes, especially for the CSF flavour (around or less 50 per cent).
In our sample, this parameter and the centroid shift are the most efficient as found also in
\citet{Lovisari:x_ray_morphology}.
%\textcolor{red}{%
The asymmetry parameter, with a superimposition percentage around 55 per cent is the third most efficient parameter.
%}
On the other hand the third-order power ratio, largely used
on X-ray maps \citep[see e.g.][]{Rasia_et_al_2013:X_ray_Morphological_Estimators_for_Galaxy_Clusters},
shows a wide overlap between the two populations (around 65 percent, reaching the extreme of 75 per cent).
The different response of the power ratio on SZ maps with respect to X-ray maps is mostly caused by
its strong dependence on the signal-to-noise ratio \citep{Poole_et_at_2006:The_impact_of_mergers_on_relaxed_X-ray_clusters_[centroid_shift]}
or gradient of the signal. In SZ, it is particularly affected by the instrumental beam \citep[see e.g.][]{Donahue_et_al_2016}.
Since we have not accounted for all these aspects, that may strong depend on the detection instrument, our conclusions regarding $ P_3/P_0 $ should not still be applicable to observational results.
%}
%

\begin{figure}
	\begin{minipage}{0.5\columnwidth}
		\hspace{-0.5cm}
		\includegraphics[width=1.1\columnwidth]{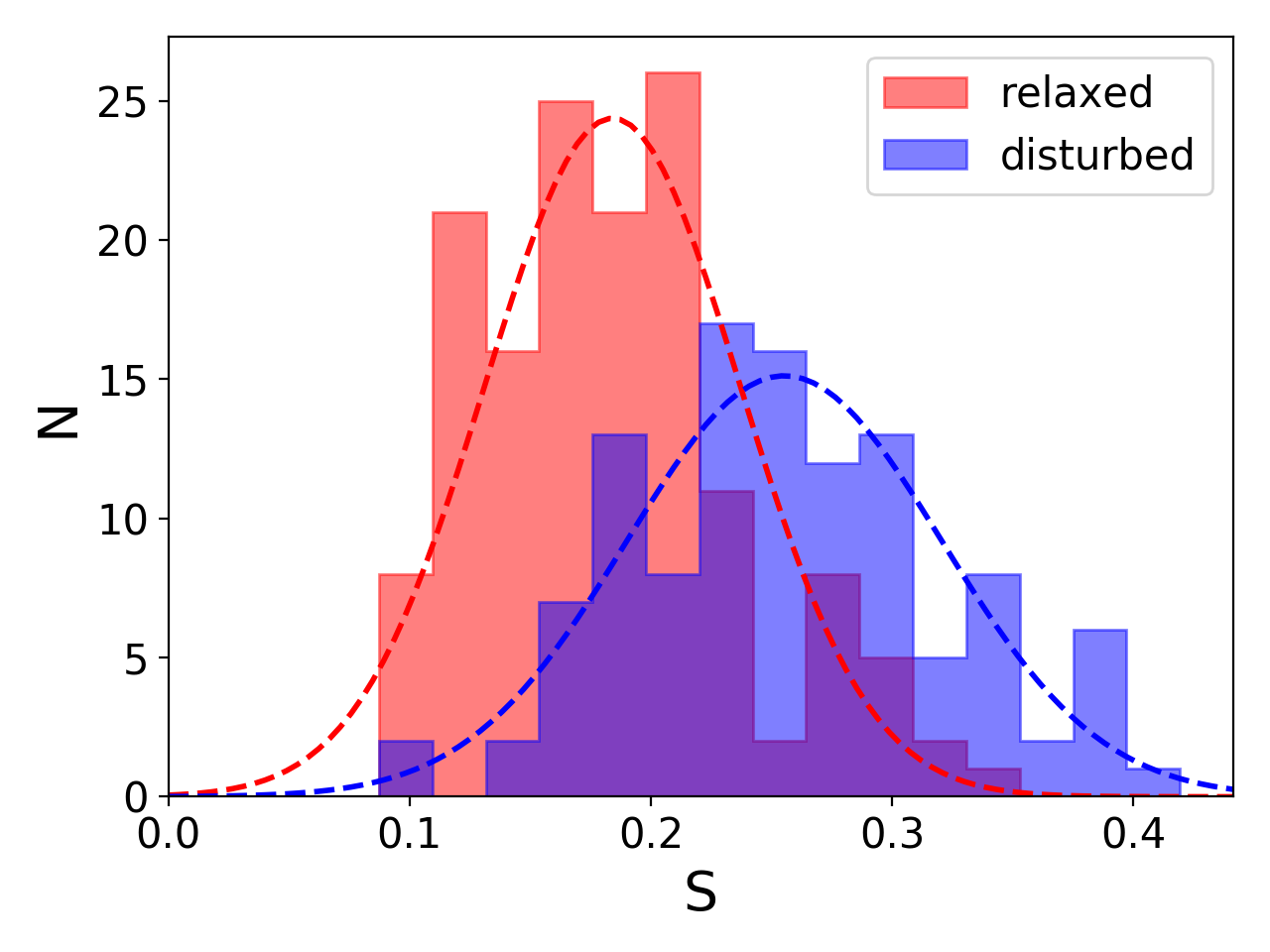}
		
	\end{minipage}% 
	\begin{minipage}{0.5\columnwidth}
		\hspace{-0.1cm}
		\includegraphics[width=1.1\columnwidth]{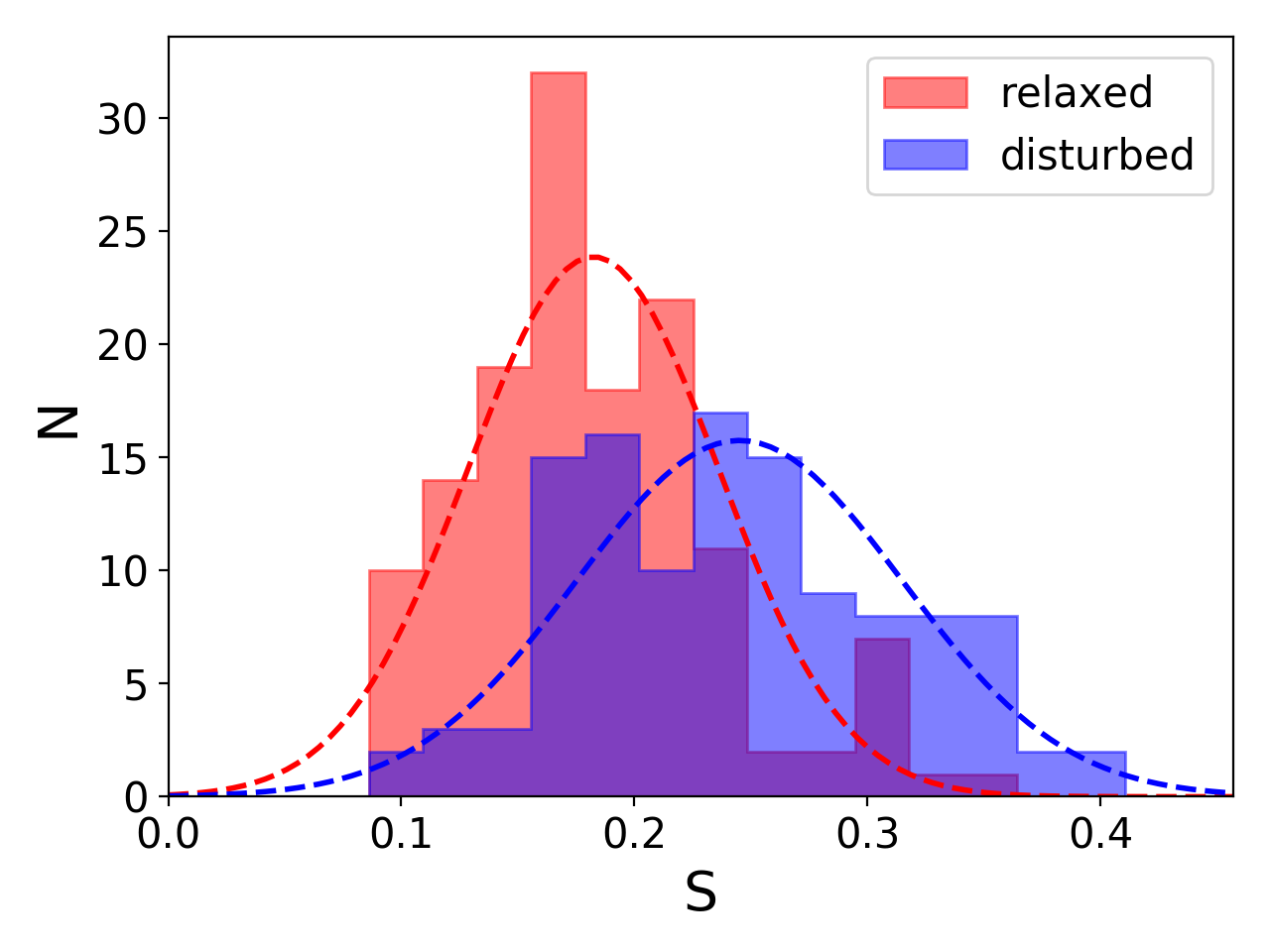}
		
	\end{minipage} 
	\begin{minipage}{0.5\columnwidth}
		\hspace{-0.5cm}
		\includegraphics[width=1.1\columnwidth]{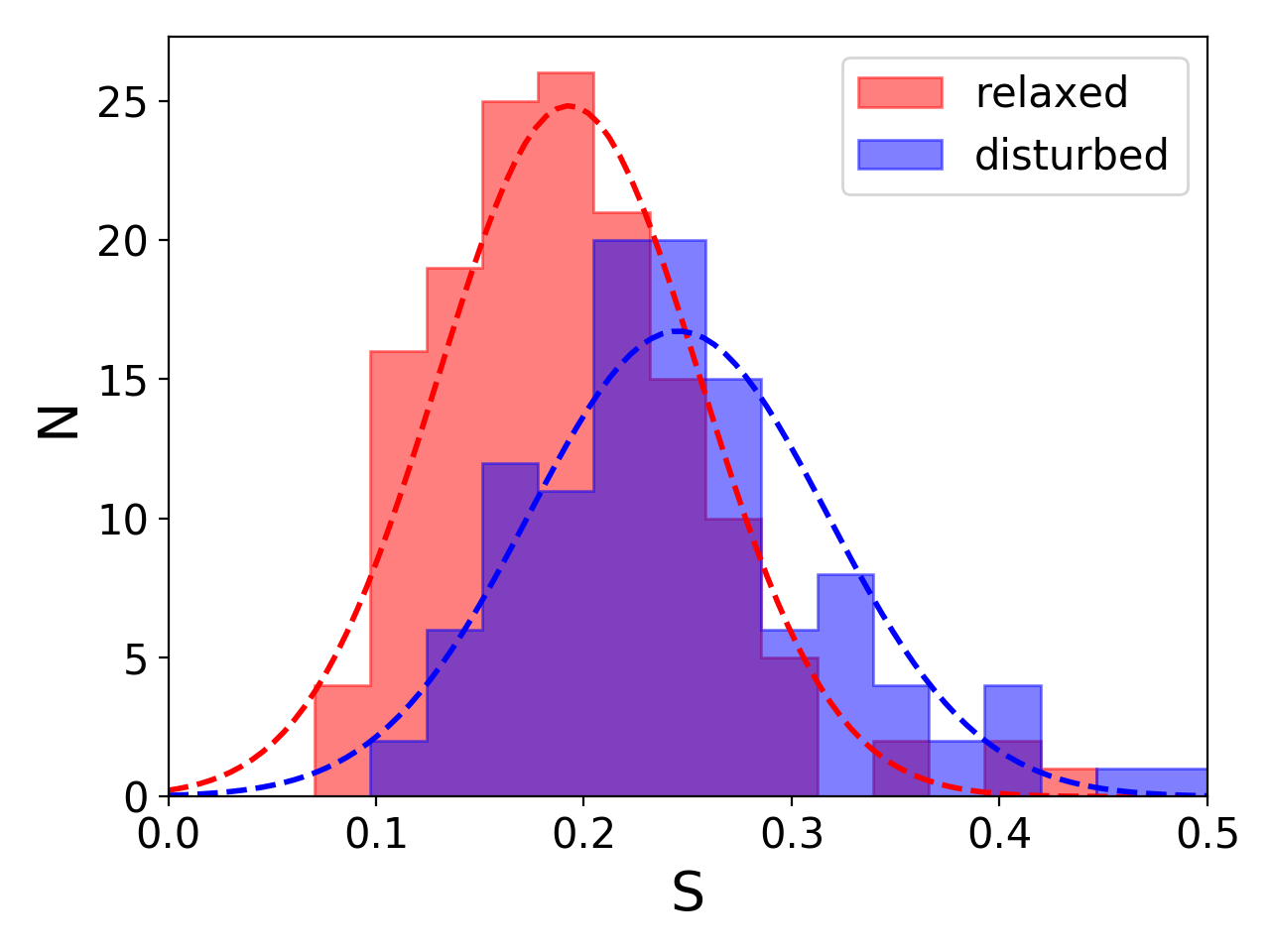}
		
	\end{minipage}%
	\begin{minipage}{0.5\columnwidth}
		\hspace{-0.1cm}
		\includegraphics[width=1.1\columnwidth]{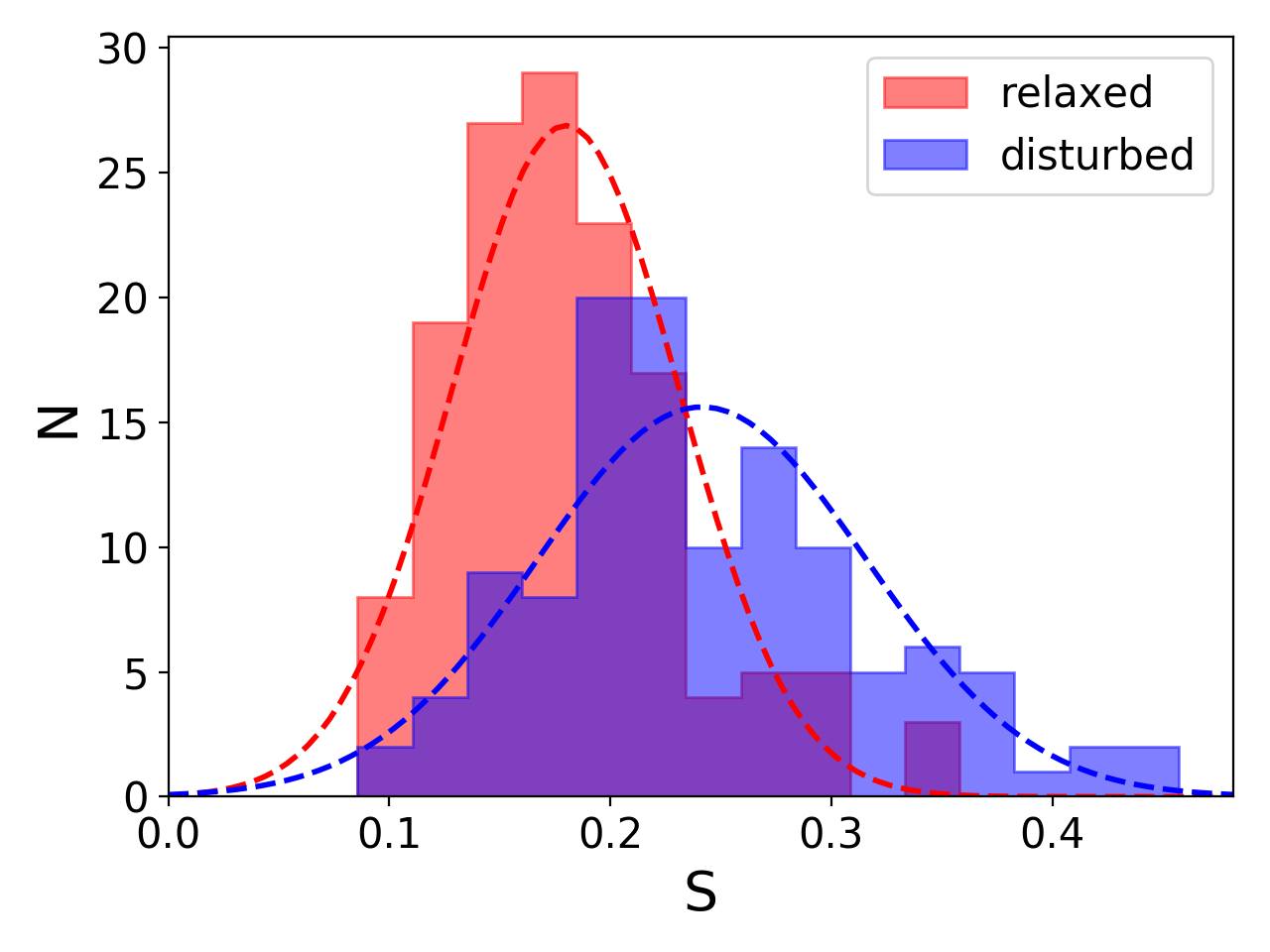}
		
	\end{minipage}
	\caption{Distribution of $S$ computed inside $R_{ap}=R_{vir}$ at the four different redshifts (0.43 top left, 
	0.54 top right, 0.67 bottom left and 0.82 bottom right) and for the CSF flavour. The red and blue bars indicate the 
	relaxed and unrelaxed populations respectively, from the a priori discrimination through the 3D indicators. The dashed lines refers respectively to the relaxed (red) and disturbed (blue) populations gaussian fit.}
	\label{fig: S parameter distributions with pre-screening under best conditions at all redshifts}   
\end{figure}

\begin{figure}
	\begin{minipage}{0.5\columnwidth}
		\hspace{-0.5cm}
		\includegraphics[width=1.1\columnwidth]{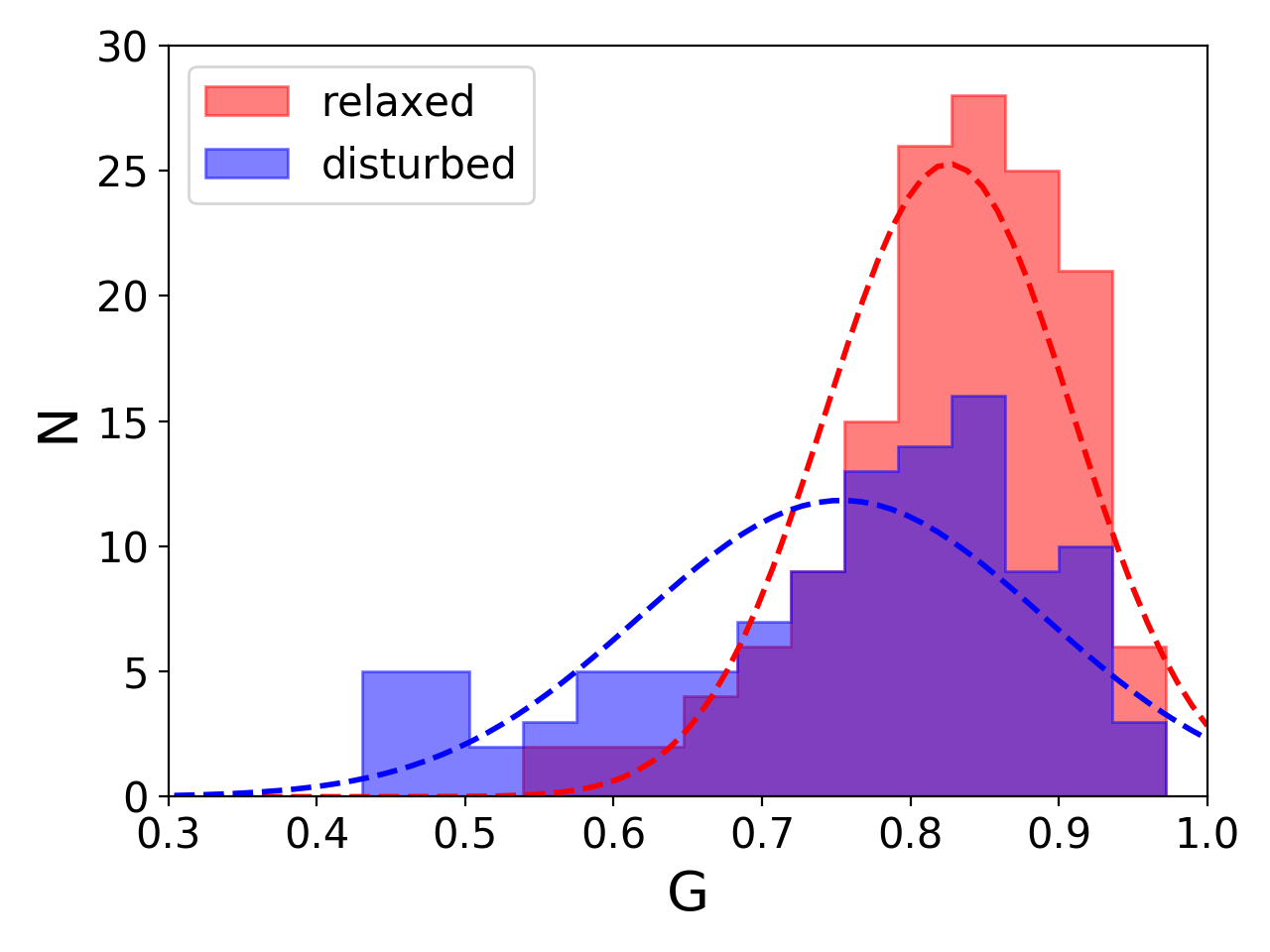}
		 
	\end{minipage}% 
	\begin{minipage}{0.5\columnwidth}
		\hspace{-0.1cm}
		\includegraphics[width=1.1\columnwidth]{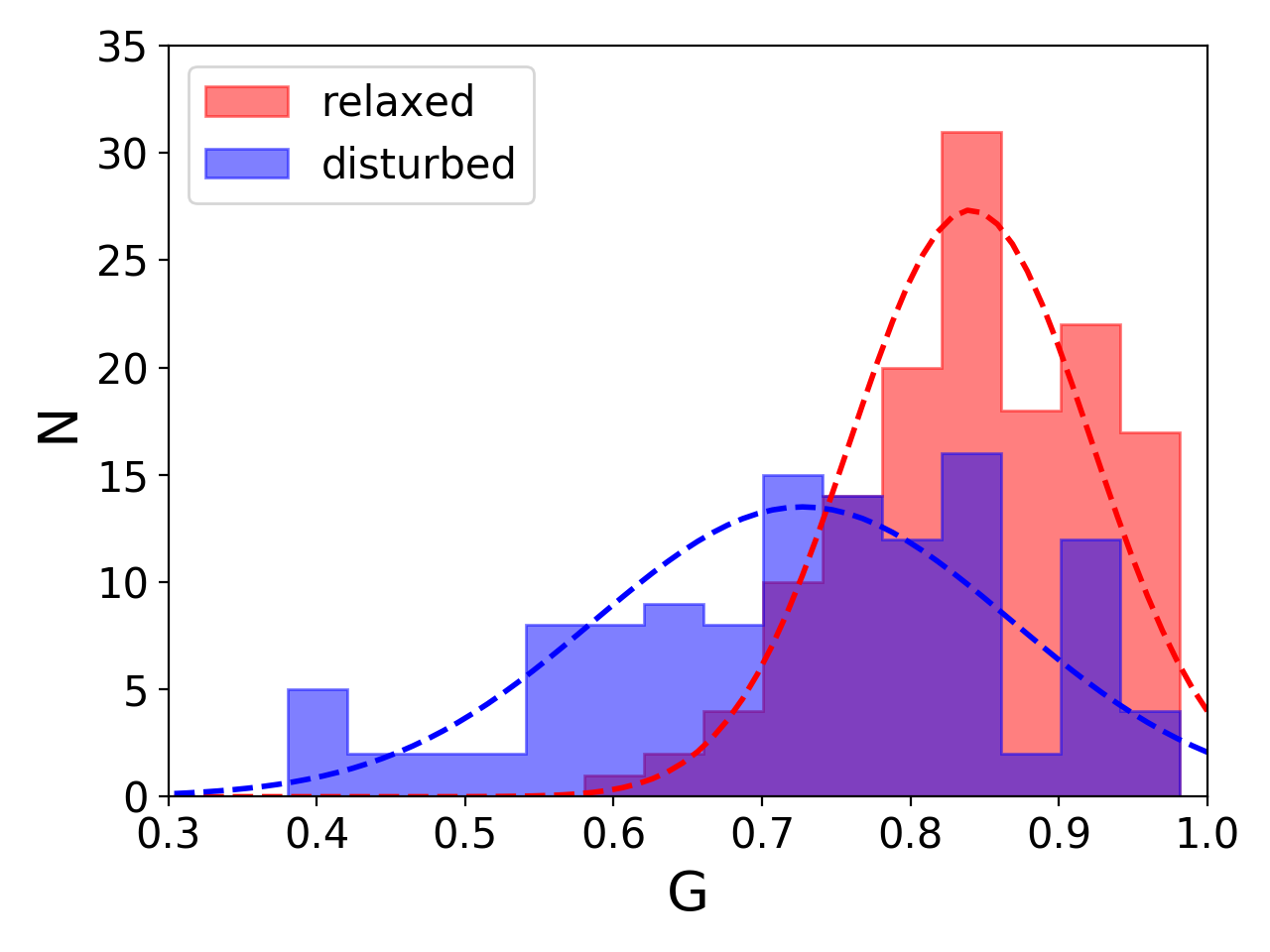}
		 
	\end{minipage} 
	\begin{minipage}{0.5\columnwidth}
		\hspace{-0.5cm}
		\includegraphics[width=1.1\columnwidth]{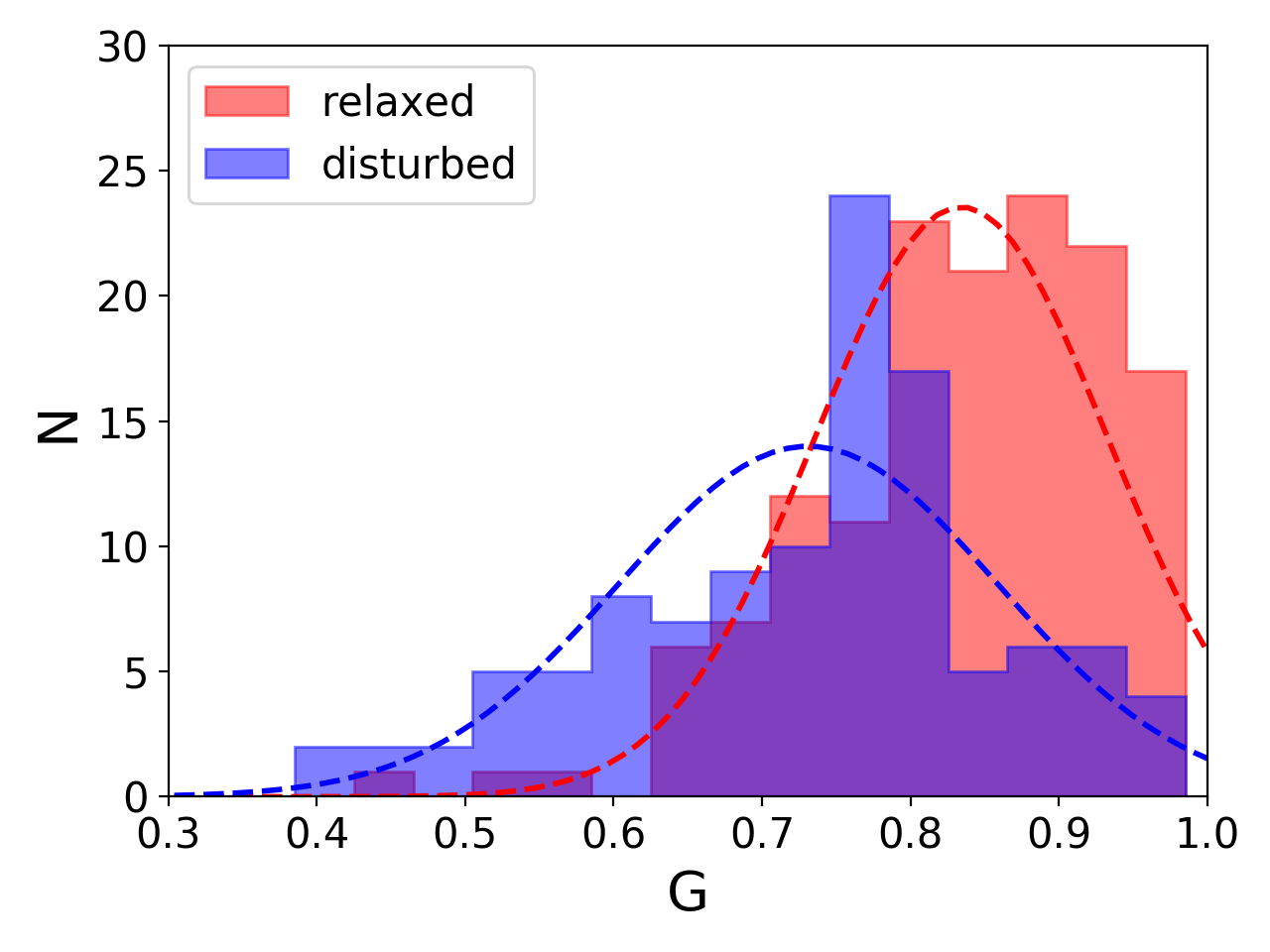}
		 
	\end{minipage}%
	\begin{minipage}{0.5\columnwidth}
		\hspace{-0.1cm}
		\includegraphics[width=1.1\columnwidth]{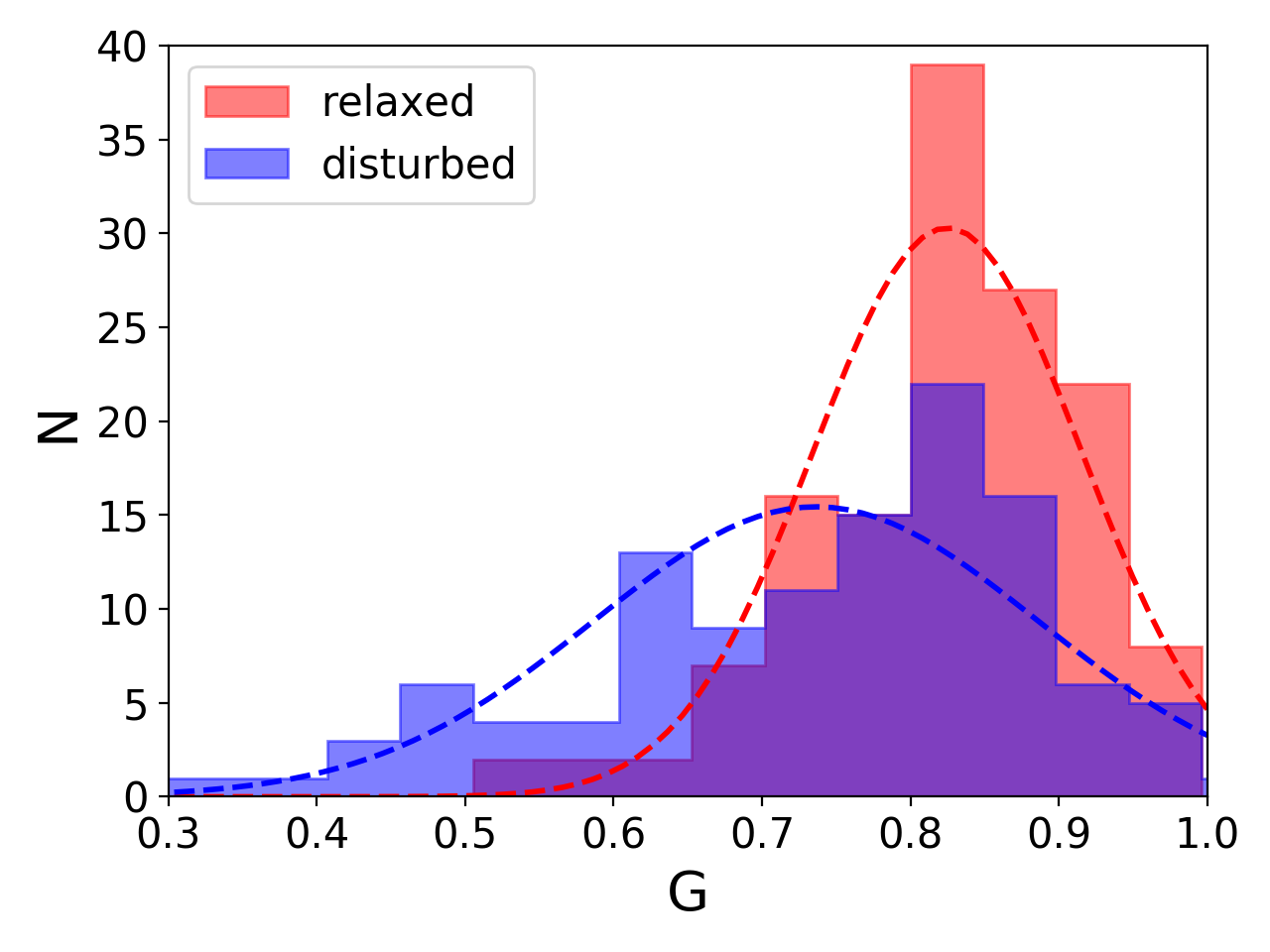}
		 
	\end{minipage}
	\caption{Distribution of the $G$ parameter computed inside $R_{ap}=R_{vir}$ for the CSF flavour (redshift 0.43 
	top left, 0.54 top right, 0.67 bottom left and 0.82 bottom right). Red and blue bars indicate the relaxed and disturbed 
	populations while red and blue dashed lines refers respectively to the gaussian fit on relaxed and disturbed populations.}
	\label{fig: G parameter distributions with pre-screening under best conditions at all redshifts}   
\end{figure}

Among the two new parameters we have previously introduced, the strip parameter performs better than the Gaussian
fitting parameter. This is also evident by comparing %
Fig.~\ref{fig: S parameter distributions with pre-screening under best conditions at all redshifts}
and~\ref{fig: G parameter distributions with pre-screening under best conditions at all redshifts} where the large overlapping area
or the two populations in the $ G $ histogram is remarkable. The inefficiency of this parameter is mostly due to the fact that the Gaussian fitting procedure smooths 
and reduces the impact of small substructures and that $ G $ is largely affected by projection effects. Indeed,
a dynamically disturbed cluster may appear regular when smoothed and observed from
a particular line of sight. This effect leads to the presence of a cospicuous fraction of disturbed clusters 
identified as relaxed.
%
% VERSIONE CON MULTIROW: FUNZIONA SOLTANTO AGGIUNGENDO IL PACCHETTO MULTIROW, linee 52-110
\begin{table}
	\caption{Results from the application of the $A$, $c$, $\log_{10}(P_3/P_0)$, $w$, $G$ and $S$ parameter to the SZ 
		maps of the sample, for the four analysed
		redshifts and the two flavours of the simulation. $ R_{ap} $ is the best aperture radius as derived
		from the minimum KS test probability, $ p_m $. We indicate with $ s_p $ the
		superimposition percentage of the distributions of the parameters for the two populations.}
	\label{tab:superimposition_percentages_allparams}
	\centering
	\begin{tabular}{cccccc}
		\hline
		& & \multicolumn{2}{c}{CSF} & \multicolumn{2}{c}{NR}\\
		$ R_{ap} \ (R_{vir})$ & $ z $ & $s_p$ & $p_m$ & $s_p$ & $p_m$\\
		\hline
		\multicolumn{6}{c}{$ A $}\\
		\hline
		\multirow{4}{*}{$ 0.75$} & 0.43 & 55\% & 8.7e-13 & 59\% & 6.7e-10 \\
		& 0.54 & 50\% & 8.5e-17 & 49\% & 1.4e-14 \\
		& 0.67 & 58\% & 1.8e-11 & 63\% & 1.2e-09 \\
		& 0.82 & 61\% & 9.2e-12 & 53\% & 6.8e-13\\
		\hline
		\multicolumn{6}{c}{$ c $}\\
		\hline
		\multirow{4}{*}{$ 0.25$} & 0.43 & 51\% & 4.8e-14 & 57\% & 2.3e-13\\
		& 0.54 & 47\% & 2.0e-16 & 53\% & 5.9e-13\\
		& 0.67 & 47\% & 6.8e-17 & 52\% & 3.8e-16\\
		& 0.82 & 48\% & 4.2e-16 & 57\% & 3.0e-12\\
		\hline
		\multicolumn{6}{c}{$ \log_{10}(P_3/P_0) $}\\
		\hline
		\multirow{4}{*}{$1.00$} & 0.43 & 65\% & 3.8e-06 & 69\% & 1.4e-05 \\
		&0.54 & 75\% & 8.5e-03 & 72\% & 4.0e-03 \\
		&0.67 & 62\% & 3.5e-09 & 65\% & 2.7e-08 \\
		&0.82 & 64\% & 4.8e-09 & 67\% & 2.6e-06 \\
		\hline
		\multicolumn{6}{c}{$ w $}\\
		\hline
		\multirow{4}{*}{$1.00$}& 0.43 & 53\% & 5.0e-11 & 53\% & 2.0e-11 \\
		& 0.54 & 57\% & 3.5e-11 & 46\% & 5.9e-18 \\
		& 0.67 & 54\% & 1.0e-11 & 59\% & 7.0e-11 \\
		& 0.82 & 55\% & 9.7e-13 & 55\% & 1.5e-10 \\
		\hline
		\multicolumn{6}{c}{$ S $}\\
		\hline
		\multirow{4}{*}{$1.00$} & 0.43 & 47\% & 1.3e-16 & 52\% & 1.4e-12 \\
		& 0.54 & 57\% & 1.5e-11 & 66\% & 4.3e-09 \\
		& 0.67 & 66\% & 3.1e-08 & 63\% & 2.7e-08 \\
		& 0.82 & 60\% & 1.5e-10 & 62\% & 4.2e-10 \\
		\hline
		\multicolumn{6}{c}{$ G $}\\
		\hline
		\multirow{4}{*}{1.00}& 0.43 & 70\% & 1.4e-04 & 73\% & 3.5e-03 \\
		& 0.54 & 60\% & 5.7e-10 & 71\% & 7.4e-05 \\
		& 0.67 & 58\% & 2.6e-09 & 72\% & 5.9e-04 \\
		& 0.82 & 68\% & 1.3e-05 & 67\% & 3.9e-05 \\
		\hline
	\end{tabular}
\end{table}
%
%%%%%%%%%%%%%%%%%%%%%%%%%%%%
\subsubsection{Comparison with X-ray results}
\label{subsubsec:SZvsXrays}
%We have tested the consistency of some 
We compare our morphological parameters, $ c $, $ w $ and $ \log(P_3/P_0) $, 
%by evaluating the
%correlation between our values and the ones derived from the 
with those measured in the mock X-ray \textsl{Chandra}-like maps described in section~\ref{sec:SZ and X maps}.
%\textcolor{red}{%
%Among all the parameters listed in section~\ref{sec:morphological_parameters}, only $ c $, $ w $ and $ \log(P_3/P_0) $ have been computed also from X-ray maps.
%}
To implement a fair comparison, we re-compute the SZ morphological parameter on the same sub-sample of the MUSIC clusters used for the X-ray analysis.
The sample include clusters from the radiative data set at redshifts $ z=0.43 $ and $ z=0.67 $ observed from 
three different lines of sight. We also set the same aperture radii equal to 500 kpc. Finally, for the light concentration ratio $ c $ we use an inner radius of 100 kpc.

\begin{figure}
	\includegraphics[width=1\columnwidth]{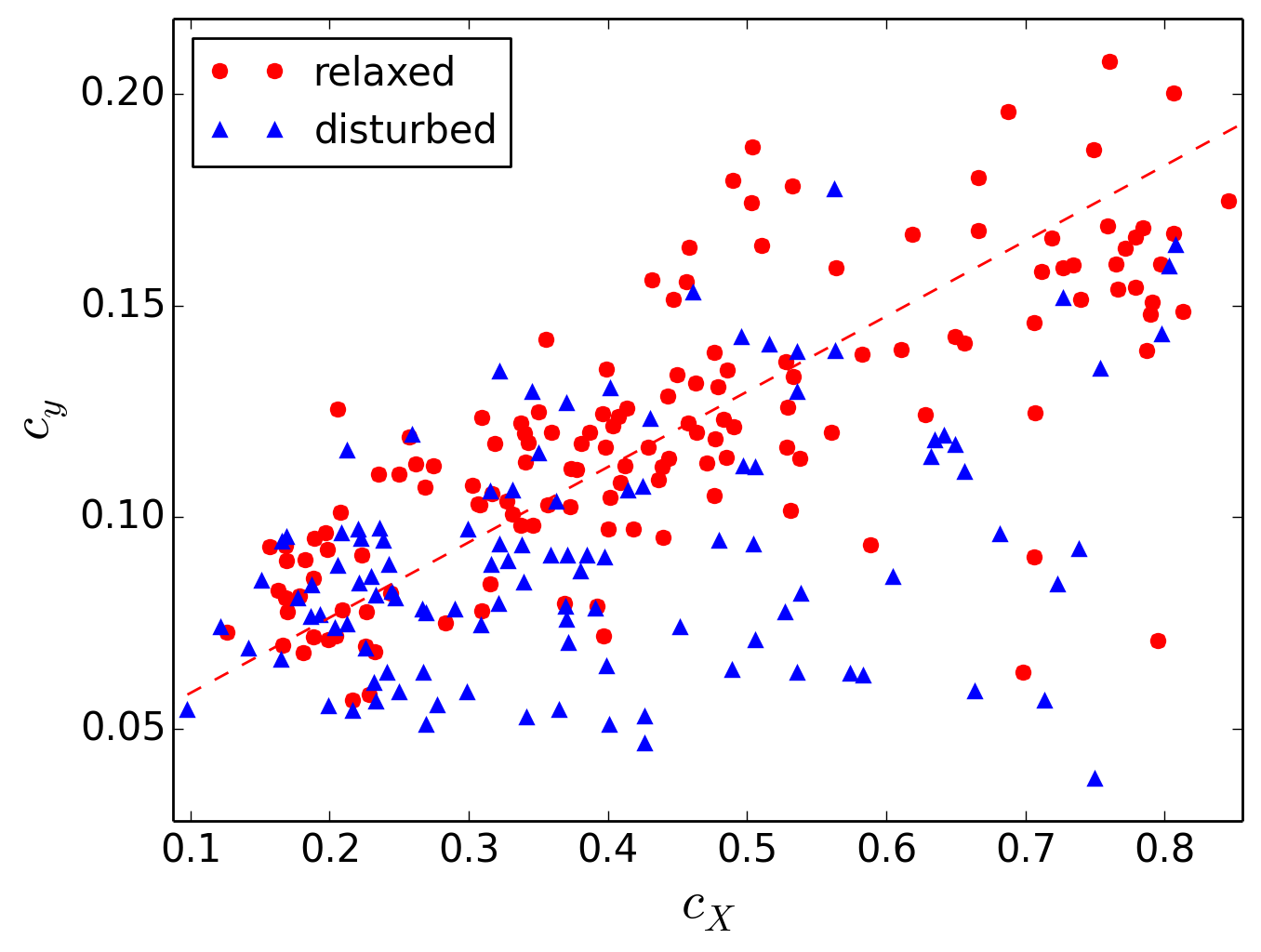}
	\caption{Scatter plot of $c$ parameter computed on X-ray ($c_X$) and SZ maps ($c_y$) for the NR clusters at $z = 0.67$.
		Relaxed 
		clusters are marked as red circles, disturbed ones are marked as blue triangles. The red dashed line shows the best 
		robust fit for the relaxed clusters only.}
	\label{fig: x ray correlation light concentration ratio same aperture radius}   
\end{figure}

Table~\ref{tab:0.67 x ray correlation coefficients same aperture radius} reports the Pearson correlation coefficients
between the X-ray and the SZ results, computed for the three parameters on three samples: one considering all the clusters, and two
sub-sets built respectively on the relaxed and disturbed populations.
%\textcolor{red}{%
%We consider that SZ and X-ray parameters show the same discrimination capability when the Pearson coefficient is larger than 0.3.
The correlation is high only for the light concentration ratio, whose corresponding scatter plot is shown in Fig.~\ref{fig: x ray correlation light concentration ratio same aperture radius}.
In addition to the correlation, one can notice the significant difference in the dynamic range of the X-ray and SZ parameters cause by the different dependence of the two signals on the electron density which is more peaked in X-ray.
%}
As expected, the two power ratios poorly correlate given the opposite performances
of this parameter in the two bands.
%\textcolor{red}{%
No significant correlation is present also in the case of the centroid shift,
probably because of the SZ and X-ray dependence on the peaked density of possible sub-structures.
%}
%\textcolor{red}{%
A similar comparison was performed in \citet{Donahue_et_al_2016}, based on the application of the parameters on maps of the CLASH sample
\citep[see section~\ref{sec:introduction} and][]{CLASH:presentation}, but applying in the SZ maps the same centre position and outer radii used for the X-ray analysis. Also in this work the concentration parameters
show a good correlation despite their different dynamic ranges.
%}
%
\begin{table}
	\centering
	\caption{Pearson correlation coefficients (CC) between the X-ray and SZ results for the three compared parameters
		$c$, $w$ and $\log(P_3/P_0)$ at redshifts 0.43 and 0.67.}
	\label{tab:0.67 x ray correlation coefficients same aperture radius}
	\begin{tabular}{lccc} 
		\hline
		\multicolumn{4}{c}{$z=0.43$} \\
		parameter & all & relaxed & disturbed \\
		\hline
		$c$ & 0.45 & 0.73 & 0.18 \\
		$\log_{10}(P_3/P_0)$ & 0.14 & 0.38 & 0.12 \\
		$w$ & 0.02  & 0.23 & -0.31 \\
		\hline
		\multicolumn{4}{c}{$z=0.67$} \\
		parameter & all  & relaxed & disturbed  \\
		\hline
		$c$ & 0.61 & 0.73 & 0.41 \\
		$\log_{10}(P_3/P_0)$ & 0.25 & 0.29 & 0.09 \\
		$w$ & 0.05  & 0.46 & -0.30 \\
		\hline
	\end{tabular}
\end{table}
%
%
%%%%%%%%%%%%%%%%
\subsection{Application of the combined parameter}
\label{subsec: results M parameter}
\begin{figure*}
	\begin{minipage}{1\columnwidth}
		\includegraphics[width=0.8\columnwidth]{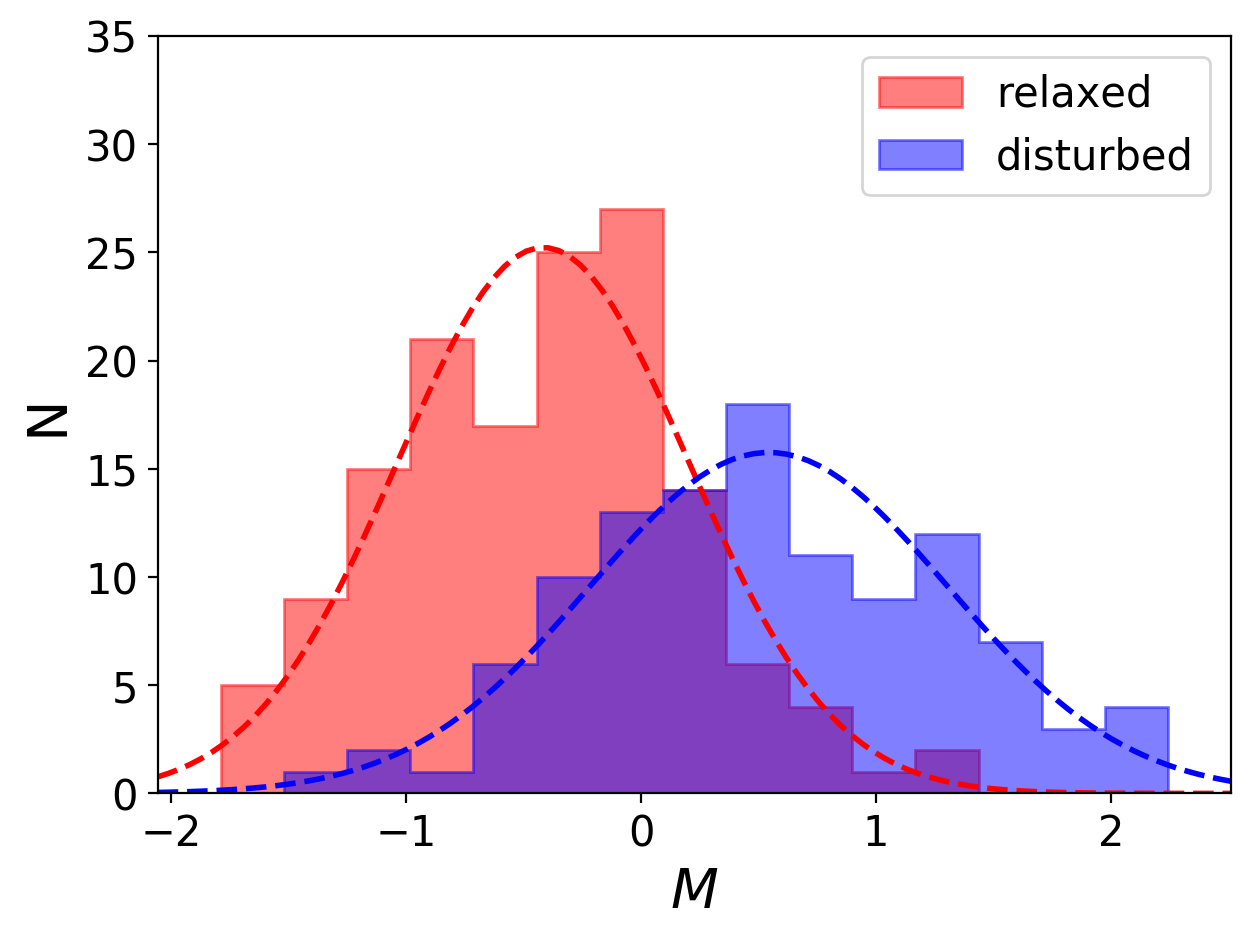}
		 
	\end{minipage}% 
	\begin{minipage}{1\columnwidth}
		\includegraphics[width=0.8\columnwidth]{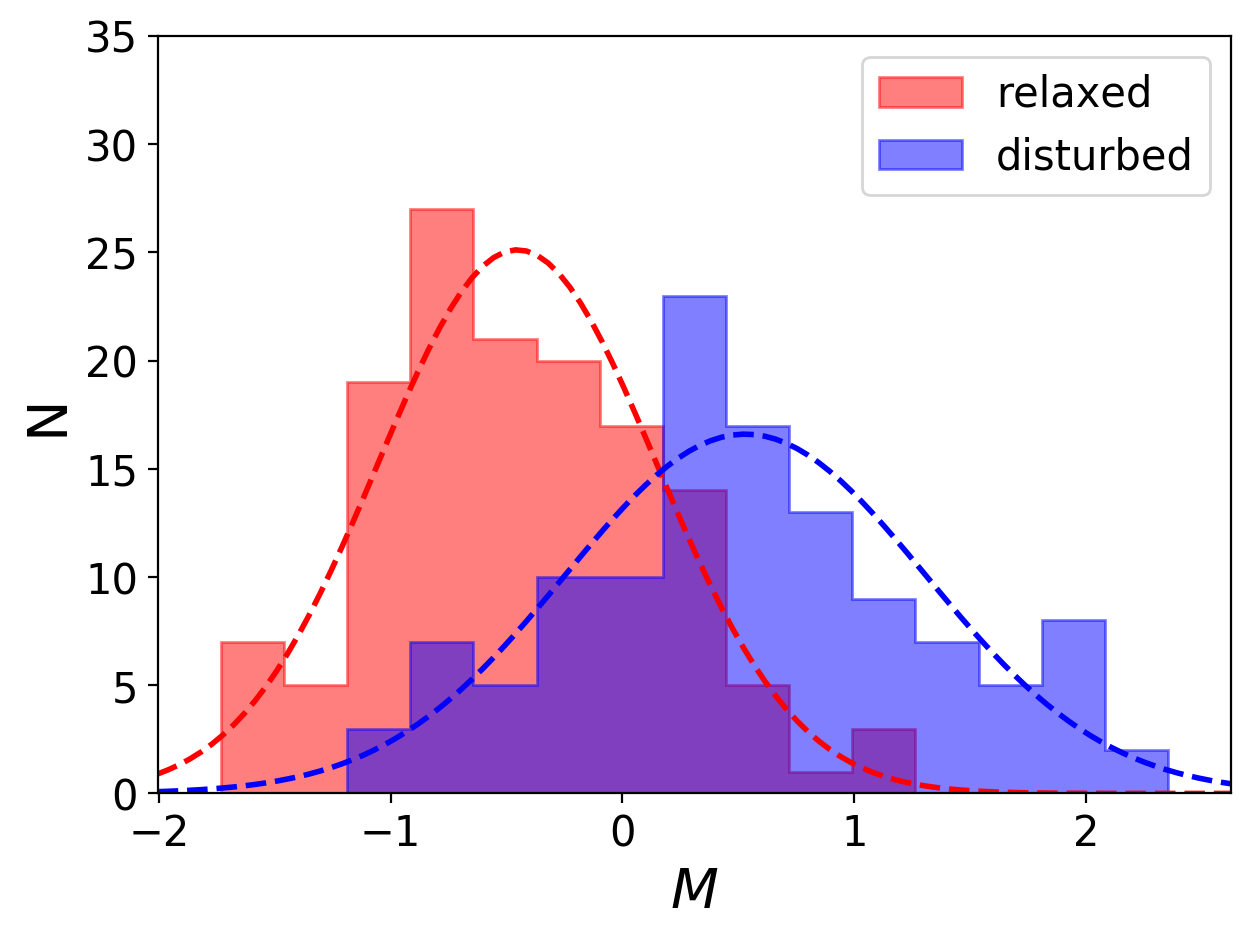}
		 
	\end{minipage} 
	\begin{minipage}{1\columnwidth}
		\includegraphics[width=0.8\columnwidth]{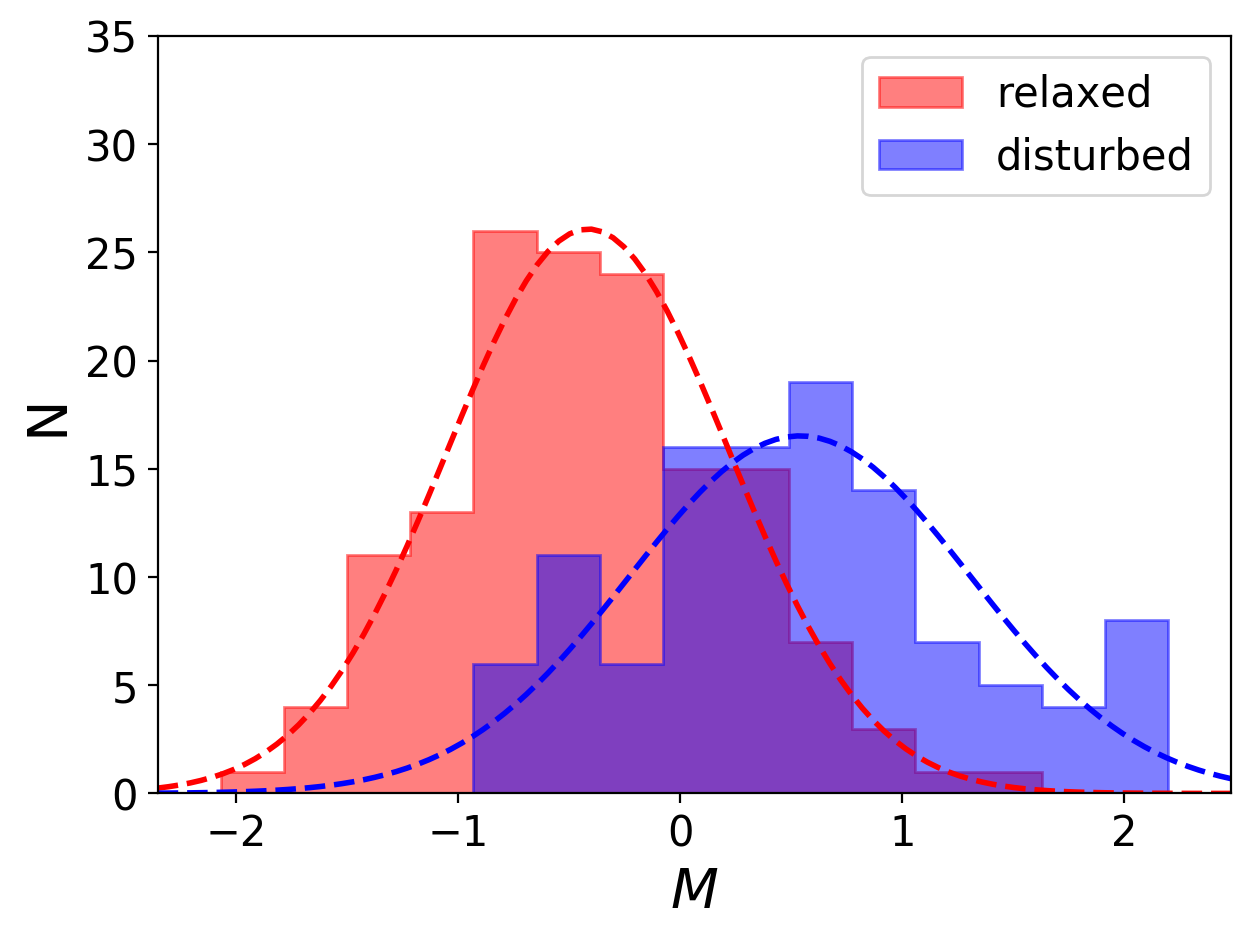}
		 
	\end{minipage}%
	\begin{minipage}{1\columnwidth}
		\includegraphics[width=0.8\columnwidth]{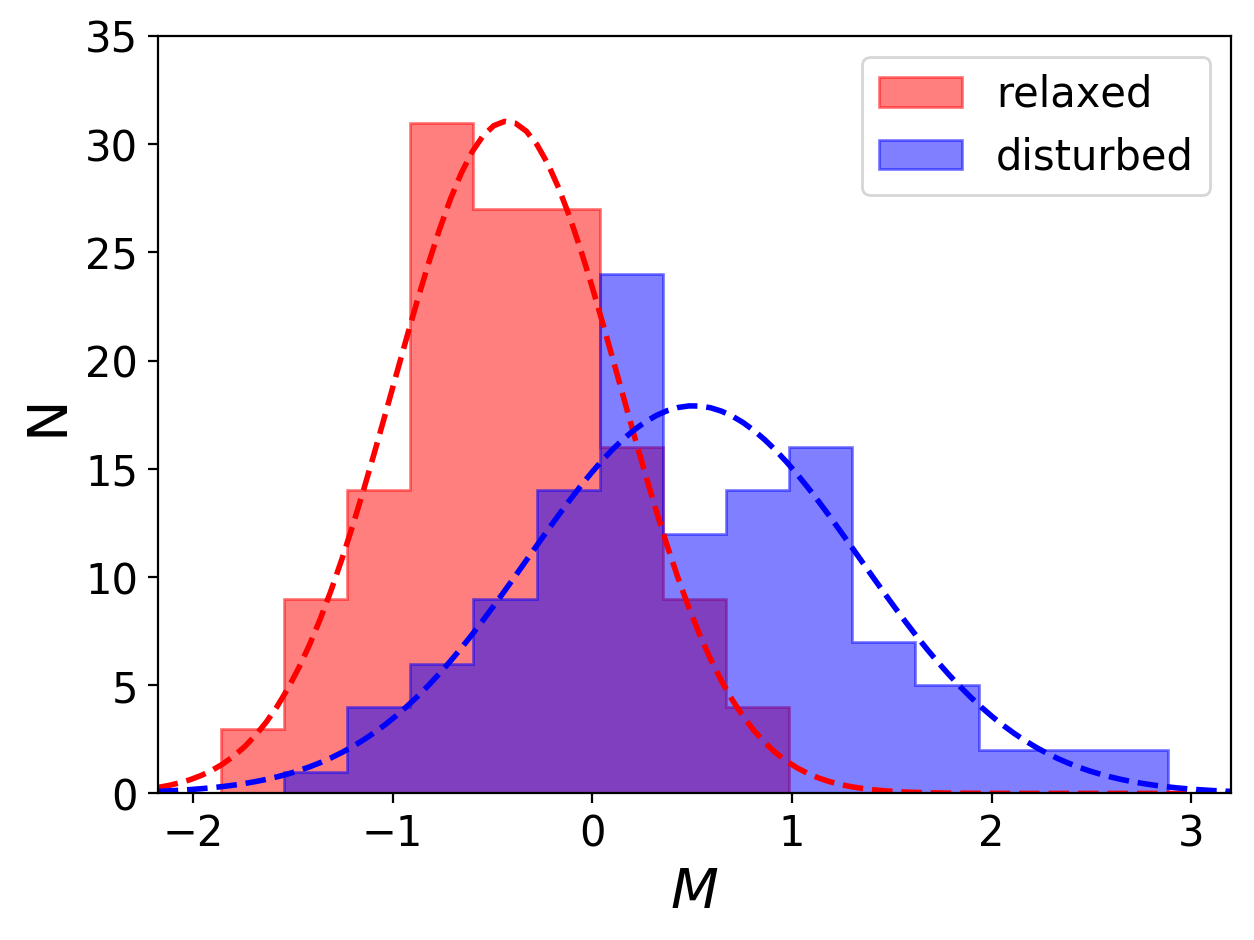}
		 
	\end{minipage}
	\caption{Distributions of $M$ computed with K-S suggested aperture radii at the four different redshifts (0.43 top left, 
	0.54 top right, 0.67 bottom left and 0.82 bottom right) for the CSF flavour. The red and blue bars indicate the relaxed 
	and unrelaxed populations respectively, determined from the 3D indicators.}
	\label{fig: M parameter distributions with pre-screening under best conditions at all redshifts}   
\end{figure*}

We compute  the combined morphological parameter $ M $, introduced in section \ref{sec: the M parameter}, following equation \ref{eq:M_parameter}, where
%\textcolor{red}{%
in particular, each $ V_i $ parameter is measured within its
most efficient aperture radius (Table~\ref{tab:superimposition_percentages_allparams}), and
its weight $ W_i $ is the absolute value of the logarithm of the corresponding 
KS probability, $ p_i $, averaged over the four redshifts and two cluster sub-sets (Table~\ref{tab:weights for M parameter}):
\begin{equation}
W_i = |\log_{10}(p_i)| \ .
\label{eq:weight general expression}
\end{equation}
With this choice the parameters showing a higher efficiency (i.e. with low values of $ p_i $) contribute
the most to $M$. 
The heaviest parameters are the light concentration ratio and the centroid shift. On the 
contrary, $ P_3/P_0 $ has the least influence. This suggests that this parameter could be neglected in real applications on 
observed SZ maps, without affecting significantly the final value of $ M $.
\begin{table}
	\centering
	\caption{Average over the redshifts and flavours of the weights for the single parameters used in the definition of the
		$M$ parameter.}
	\label{tab:weights for M parameter}
	\begin{tabular}{lc} % three columns, alignment for each
		\hline
		parameter & $W_i$  \\
		\hline
		$A$ & 9.63 \\
		$c$ & 12.31 \\
		$\log_{10}(P_3/P_0)$ & 2.80 \\
		$w$ & 10.38 \\
		$S$ & 8.10 \\
		$G$ & 3.31 \\
		\hline
	\end{tabular}
\end{table}
The distributions of $M$ for the relaxed and unrelaxed clusters are shown
in Fig.~\ref{fig: M parameter distributions with pre-screening under best conditions at all redshifts};
the corresponding overlap percentages and probabilities from the KS test are listed in
Table~\ref{tab:superimposition percentages M param}.
It is noticeable the efficiency improvement of this parameter 
%performs better in discriminating the two classes, with respect to the 
over each single parameter: the contamination level is of the order of 22 per cent for the relaxed population and
28 per cent for the disturbed population.
\begin{table}
	\centering
	\caption{Superimposition percentage $ s_p $ between the distributions of relaxed and disturbed clusters,
		and associated KS probability $ p_m $ for the $M$ parameter.}
	\label{tab:superimposition percentages M param}
	\begin{tabular}{lcccc} % three columns, alignment for each
		\hline
		$z$& \multicolumn{2}{c}{CSF} & \multicolumn{2}{c}{NR}\\
		& $ s_p $ & $ p_m $ & $ s_p $ & $ p_m $ \\
		\hline
		0.43 & 47\% & 4.2e-18 & 50\% & 3.0e-14\\
		0.54 & 45\% & 7.7e-19 & 48\% & 2.5e-16\\
		0.67 & 51\% & 1.3e-16 & 51\% & 1.2e-13\\
		0.82 & 49\% & 3.8e-16 & 51\% & 1.0e-14\\
		\hline
	\end{tabular}
\end{table}
%
%%%%%%%%%%%%%%%%%%

\subsection{Test on the stability of the parameters}
To check the stability of all parameters,
we produce SZ maps of two relaxed and
two disturbed clusters from our CSF sub-set at $ z=0.54 $ and $ z=0.82 $
 along 120 lines of sight.
 We checked that the results are similar for the
NR sub-set and at the remaining redshifts. 
%\textcolor{red}{%
Specifically, we consider lines of sight spaced in uniform steps of 9$^{\circ}$.
% for the rotation angle,
%obtaining a total of 120 different maps for each cluster.
%}
The six morphological parameters and the combined one are applied to all these maps and their distributions are drawn. The corresponding 
standard deviations are an estimate of the stability of the parameter. We list the results in Table~\ref{tab: rotations standard deviations}.
For all parameters, disturbed clusters tend to have a higher dispersion with respect to the
relaxed ones, as expected because of projection effects. The mean value of the standard deviation of $ M $, with respect to the four clusters at both redshifts, is $<{\sigma_M}> = 0.41$. 
We consider this value as the uncertainty of the $M$ parameter. 
For this reason all clusters with $M$ between -0.41 and 0.41 might
be a mixture of relaxed and disturbed objects.
%Thus it is possible to define a relaxed-disturbed hybrid population
%of clusters lying in the region centred to the threshold value of $M=0$, showing $-0.41<M<0.41$.
%
\begin{table*}
	\centering
	\caption{Standard deviations of all morphological parameters for two relaxed (\#17 and \#267)
		and two disturbed (\#44 and \#277) clusters at redshift 0.54 and 0.82, as derived from the SZ maps
		computed at 120 different lines of sight.}
	\label{tab: rotations standard deviations}
	\begin{tabular}{lcccccccc}
		\hline
		parameter & \multicolumn{4}{c}{$z=0.54$} & \multicolumn{4}{c}{$z=0.82$}\\
		& \multicolumn{2}{c}{relaxed} & \multicolumn{2}{c}{disturbed} & \multicolumn{2}{c}{relaxed} & \multicolumn{2}{c}{disturbed} \\
		& \#17 & \#267 & \#44 & \#277 &\#17 & \#267 & \#44 & \#277 \\
		\hline
		$A$ & 0.030 & 0.040 & 0.060 & 0.050 & 0.060 & 0.040 & 0.100 & 0.080 \\
		$c$ & 0.007  &  0.002 &  0.008 &  0.013 & 0.014 & 0.005 & 0.008 & 0.015 \\
		$\log_{10}(P_3/P_0)$ & 0.620 & 0.430 & 0.600 & 0.430 & 0.470 & 0.630 & 0.400 & 0.510 \\
		$w$ & 0.002 & 0.0007 & 0.004 & 0.003 & 0.001 & 0.001 & 0.005 & 0.008 \\
		$S$ & 0.030 & 0.020 &  0.060 &  0.020 & 0.040 & 0.020 & 0.070 & 0.040 \\
		$G$ & 0.070 & 0.060 & 0.050 & 0.070 & 0.040 & 0.070 & 0.070 & 0.130 \\
		\hline
		$M$ & 0.28 & 0.30 & 0.43 & 0.37 & 0.49 & 0.26 & 0.57 & 0.58 \\
		\hline
	\end{tabular}
\end{table*}
%
%\begin{table}
%	\centering
%	\caption{Standard deviations of $M$ parameter as derived from the single parameters dispersions
%		listed in Table~\ref{tab: rotations standard deviations} (see text).}
%	\label{tab: rotations renormalized standard deviations for M z0.54 and z0.82 CSF}
%	\begin{tabular}{lccccc} 
%		\hline
%		parameter & \multirow{2}{*}{$z$} & \multicolumn{4}{c}{cluster number} \\
%		&  & 17 & 267& 44 & 277 \\
%		\hline
%		\multirow{2}{*}{$M$} & 0.54 & 0.28 & 0.30 & 0.43 & 0.37 \\
%		        & 0.82 & 0.49 & 0.26 & 0.57 & 0.58 \\
%		\hline
%	\end{tabular}
%\end{table}
%
\subsection{Effects of the angular resolution}
\label{subsec:angularresolutions}
Current SZ experiments cannot reproduce the angular resolution of our maps
%as high as in hydrodynamical
%simulations 
(see section~\ref{sec:introduction}).
In this section we want to verify the impact of the resolution on the 
performance of the morphological parameters. We thus reduce the SZ signal of our maps by 
applying a Gaussian smoothing.
We choose three FWHMs similar to
the typical resolution achieved by existing telescopes in the millimetric band: 
20 arcsec for IRAM 30-m telescope, 1 arcmin for
 SPT 10-m telescope and 5 arcmin for the \textsl{Planck} telescope.
We perform the convolution only on CSF maps at $ z = 0.54 $ and $ z = 0.82 $.

The superimposition percentages of the histograms of the relaxed and unrelaxed classes for the $M$ parameter are listed in
Table~\ref{tab: superimposition percentages of M parameter with angular resolution}.
The overlap is expected to increase with the value of the FWHM (i.e. with increasing resolution),
%nevertheless we find a minimum at 1 arcmin, which, however, is about 5 per cent smaller than the superimposition
%we find at 20 arcsec.
nevertheless we find a decrease of 1 per cent and 3 per cent of the superimposition at 20 arcsec and
1 arcmin, respectively.
%\textcolor{red}{%
This may be explained by considering that the $M$ parameter depends on the overall
morphology, thus small-scale effects are negligible at resolutions of the order of few arcminutes. With respect to the individual parameters, we found that the angular resolution barely affects their ability to distinguish between the two dynamical classes when the morphological parameters are sensitive to the properties of the cluster core, such as the light concentration ratio. At the same time, the resolution has a large impact on the results derived from the parameters built to enhance the presence of substructures such as the strip parameter.
%}
%
\begin{table}
	\centering
	\caption{Superimposition percentages of $M$ with respect to the FWHM of the Gaussian filter applied
		to the maps to simulate a decreasing angular resolution.}
	\label{tab: superimposition percentages of M parameter with angular resolution}
	\begin{tabular}{ccc} 
		\hline
		FWHM (arcsec)& \multicolumn{2}{c}{$z$} \\
		& 0.54 & 0.82 \\
		\hline
		$20$ & 43\% & 48\% \\
		$60$ & 41\% & 46\% \\
		$300$ & 51\% & 54\% \\
		\hline
	\end{tabular}
\end{table}
\section{Morphology and hydrostatic mass bias}
\label{sec: mas bias correlation}
We investigate, here, the correlation between the morphological parameter $M$  and the 
deviation from the hydrostatic equilibrium in the ICM.
For each simulated cluster, we define the hydrostatic mass bias as follows: 
\begin{equation}
b_M = (M_{500,HSE} - M_{500}) / M_{500}.
\label{eq:Mass-bias definition}
\end{equation}
In the expression above, $M_{500}$ is the mass obtained by summing the gas and DM particles inside $R_{500}$. 
$M_{500,HSE}$ is the mass computed under the assumption of the hydrostatic 
equilibrium expressed as:
\begin{equation}
M_{500,HSE} = - \frac{k_BTr}{G\mu m_H} \left( \frac{d\text{ln}\rho}{d\text{ln}r} + \frac{d\text{ln}T}{d\text{ln}r} \right) \ ,
\label{eq:Mass HSE}
\end{equation}
where $k_B$ and $G$ are the Boltzmann and gravitational constants, $r$ is the radius from the centre of the cluster, $\mu$ 
is the mean molecular weight, $m_H$ the hydrogen mass, $\rho$ the density and $T$ the mass-weighted temperature 
\citep{Sembolini_et_al_2013:The_MUSIC_of_Galaxy_Clusters_I}.
The mass bias has already been analysed in many works on hydrodynamical simulations, lensing and X-ray observations like 
\citet{Kay_et_al_2004:mass_bias_MNRAS.355.1091K}, \citet{Rasia_et_al_2006_mass_bias_first_article_MNRAS.369.2013R}, 
\citet{Nagai_et_al_2007:mass_bias_ApJ...655...98N}, 
\citet{Jeltema_et_al_2008:Cluster_Structure_in_Cosmological_Simulations...[centroid_shift]}, 
\citet{Piffaretti_et_al_2008:mass_bias_A&A...491...71P}, \citet{Zhang_et_al_2010:mass_bias_ApJ...711.1033Z}, 
\citet{Meneghetti_et_al_2010:mass_bias_A&A...514A..93M}, \citet{Becker_et_al_2011:mass_bias_ApJ...740...25B} and 
\citet{Sembolini_et_al_2013:The_MUSIC_of_Galaxy_Clusters_I}. 

The sources of the asymmetry in the ICM distribution should also impact the hydrostatic equilibrium. We, therefore, investigate correlation between the absolute value 
of the mass bias $|b_M|$ and the $M$ parameter in our sample. The results are reported in Fig. \ref{fig: mass bias versus 
m parameter CSF} for CSF clusters (NR clusters have a similar behaviour). For each redshift and simulated sub-set (NR/CSF) we compute the 
Pearson correlation coefficient and report the results in Table \ref{tab:mass bias vs M, Pearson CC}. This 
coefficient is almost always below 0.30. This rather weak correlation leads us to conclude that there is no strong connection between $|b_M|$ and the morphology of the cluster as quantified by our indicators.\\ 

%%%%% UNTIL HERE%%%%

\begin{figure}
	\centering
	\includegraphics[width=1\columnwidth]{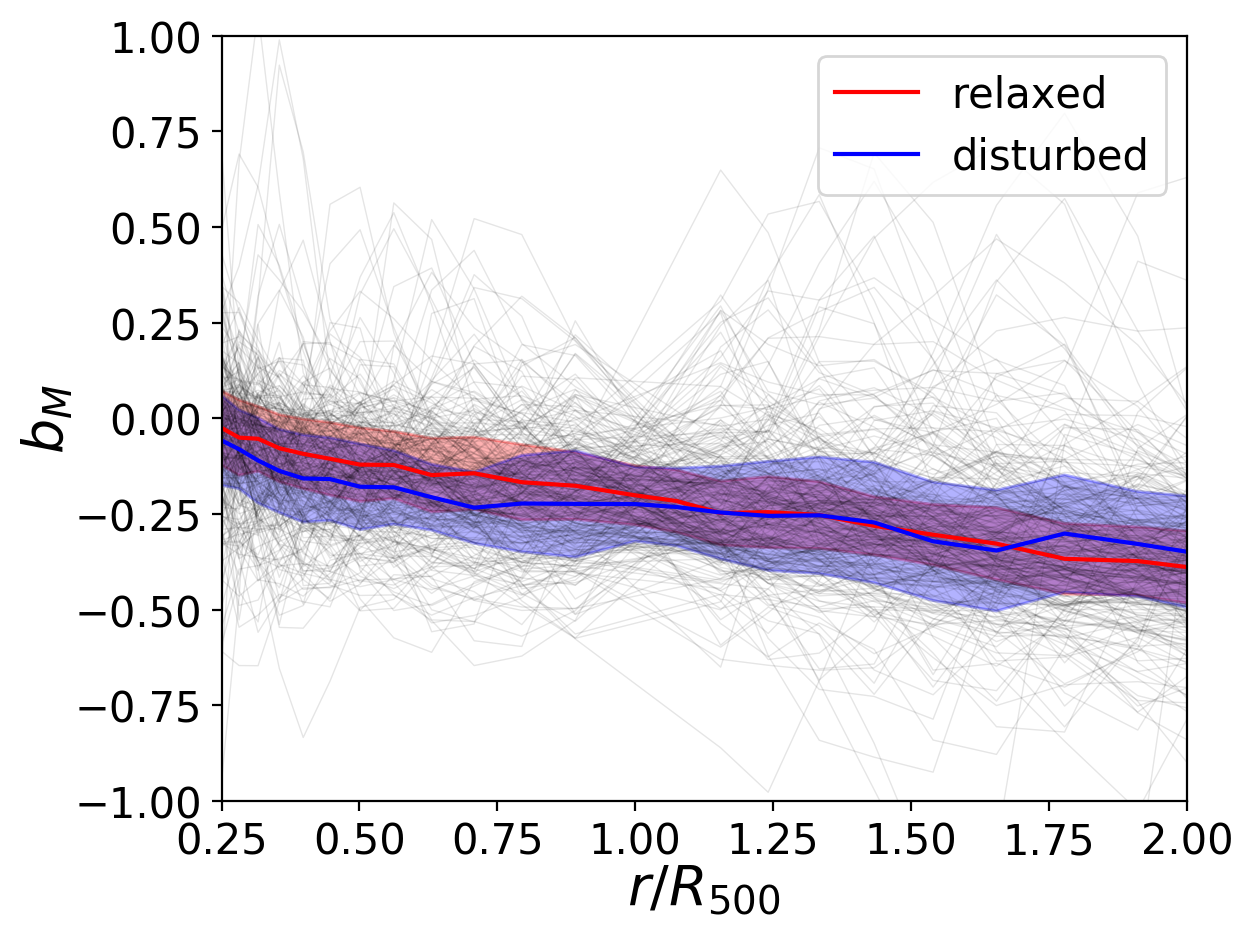}
	\caption{Radial median profiles of $b_M$  for relaxed (red) and disturbed (blue) clusters of the sample at redshift 
		$z=0.54$ and CSF flavour. Gray shaded lines represents the highly scattered single cluster profiles from which the 
		median are computed. The coloured regions indicate the median absolute deviations of the median profiles.}
	\label{fig: HMB median profile}   
\end{figure}

This result suggest that the amplitude of the mass bias is not tightly connected to the dynamical state of a cluster. However, It is worth to stress that the mass bias values we have used for this comparison are computed within $R_{500}$. This may be a 
limiting factor, since a different radial value could be used. For this reason we compute the median profiles of the mass 
bias along the cluster radius, for the relaxed and disturbed clusters segregated according to the 3D parameters. We 
show these profiles in Fig. \ref{fig: HMB median profile}, where a consistent superimposition between the two 
populations within the median absolute deviation can be seen from 0.8 $R_{500}$ and beyond.
It can be also seen that the scatter of 
the median profile for the disturbed clusters is higher, since they are expected to show a more significant deviation from 
the hydrostatic equilibrium with respect to the relaxed ones. Among the others, 
\citet{Biffi_et_al_2016:On_the_Nature_of_Hydrostatic_Equilibrium_in_Galaxy_Clusters} already investigated the median 
radial profile of the mass bias, using a small sample of 6 relaxed and 8 disturbed clusters from a simulation which also 
includes feedback from AGN. They find that the radial median profiles of the relaxed and disturbed clusters have very 
little differences, as in our case, and that the mass bias is slightly lower (in absolute value) for the relaxed clusters.
\begin{figure*}
	\centering
	\subfloat[]
	{\includegraphics[width=1\columnwidth]{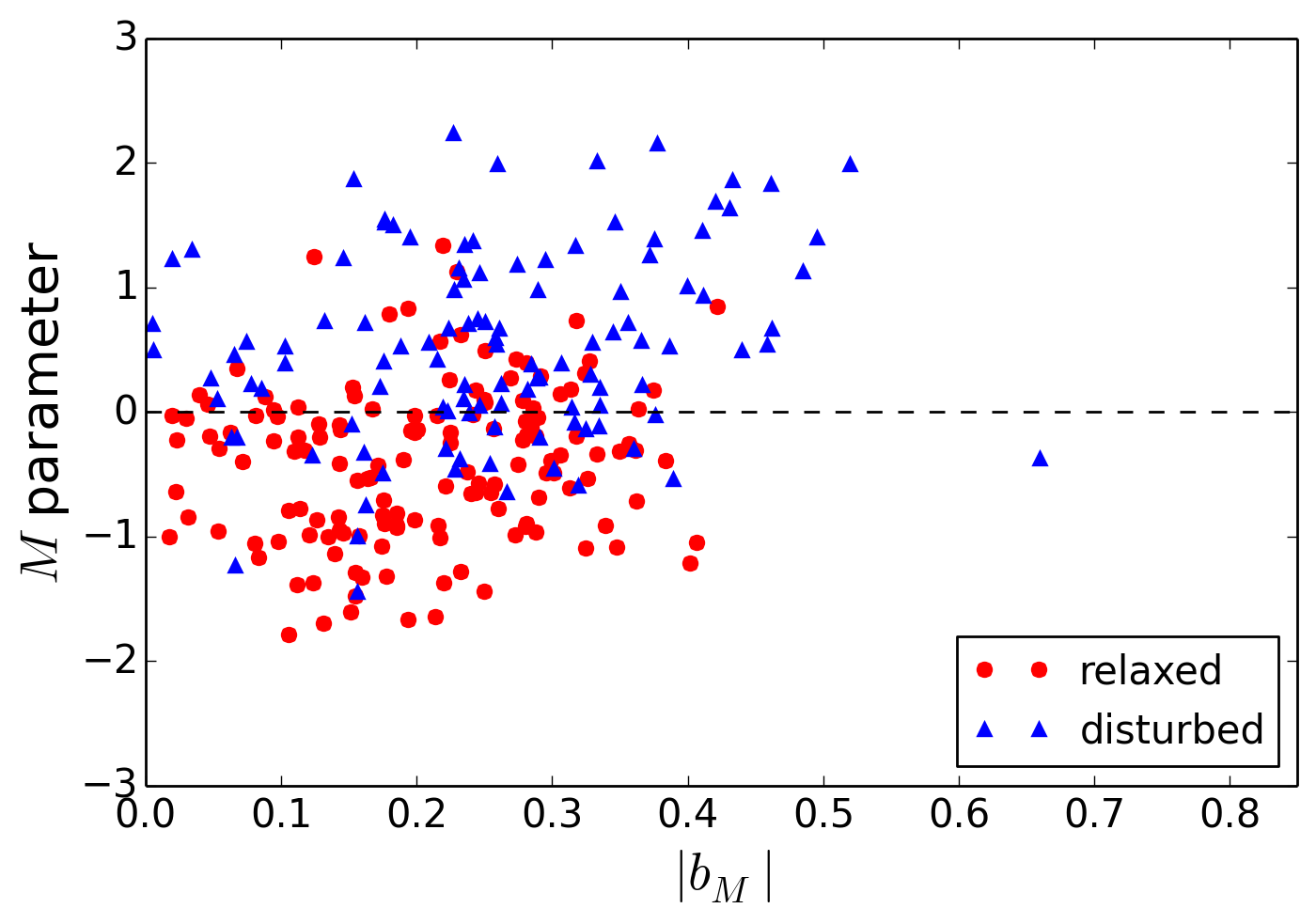}}
	\hspace{5mm}
	\subfloat[]
	{\includegraphics[width=1\columnwidth]{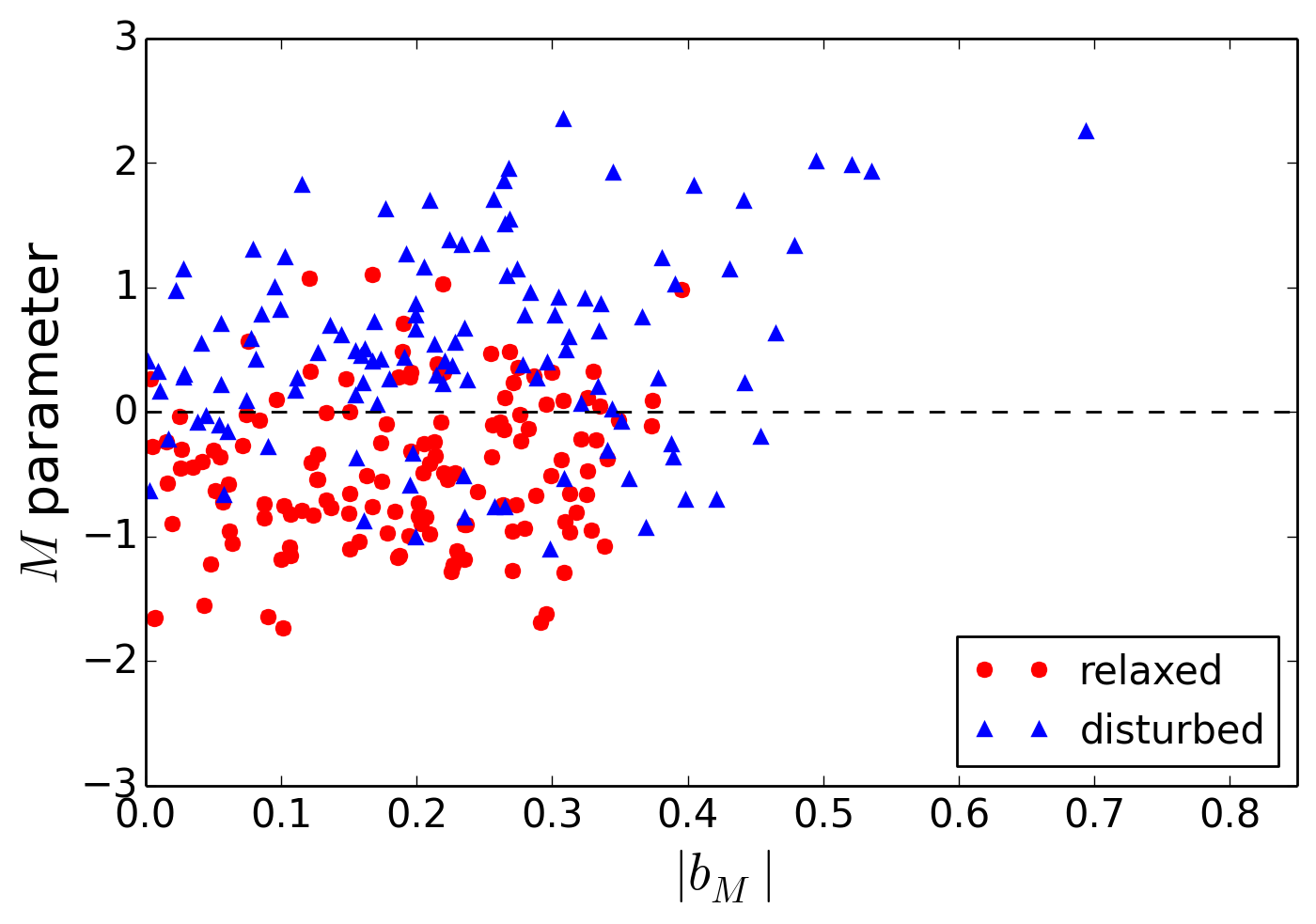}}
	\hspace{5mm}
	\subfloat[]
	{\includegraphics[width=1\columnwidth]{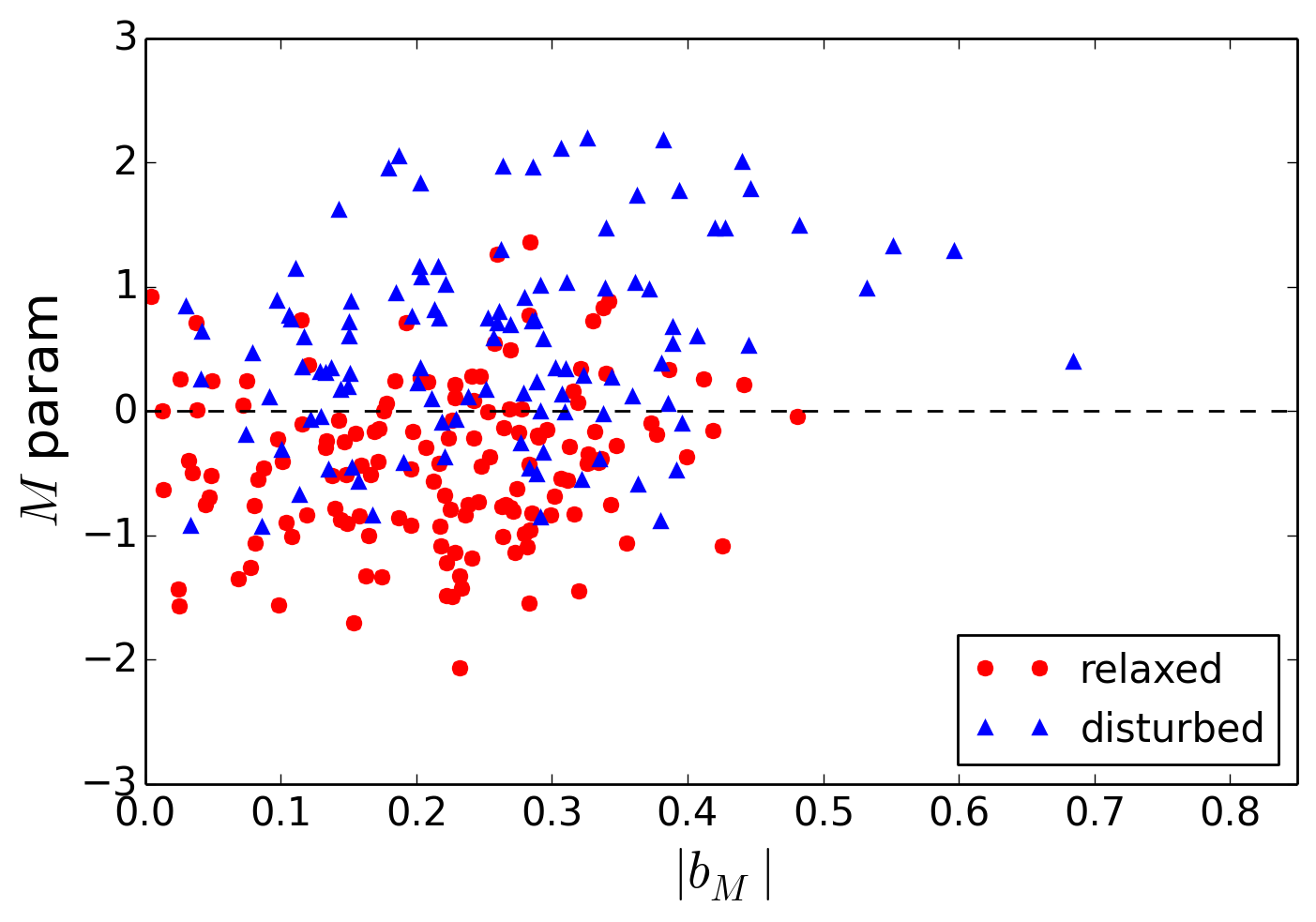}}
	\hspace{5mm}
	\subfloat[]
	{\includegraphics[width=1\columnwidth]{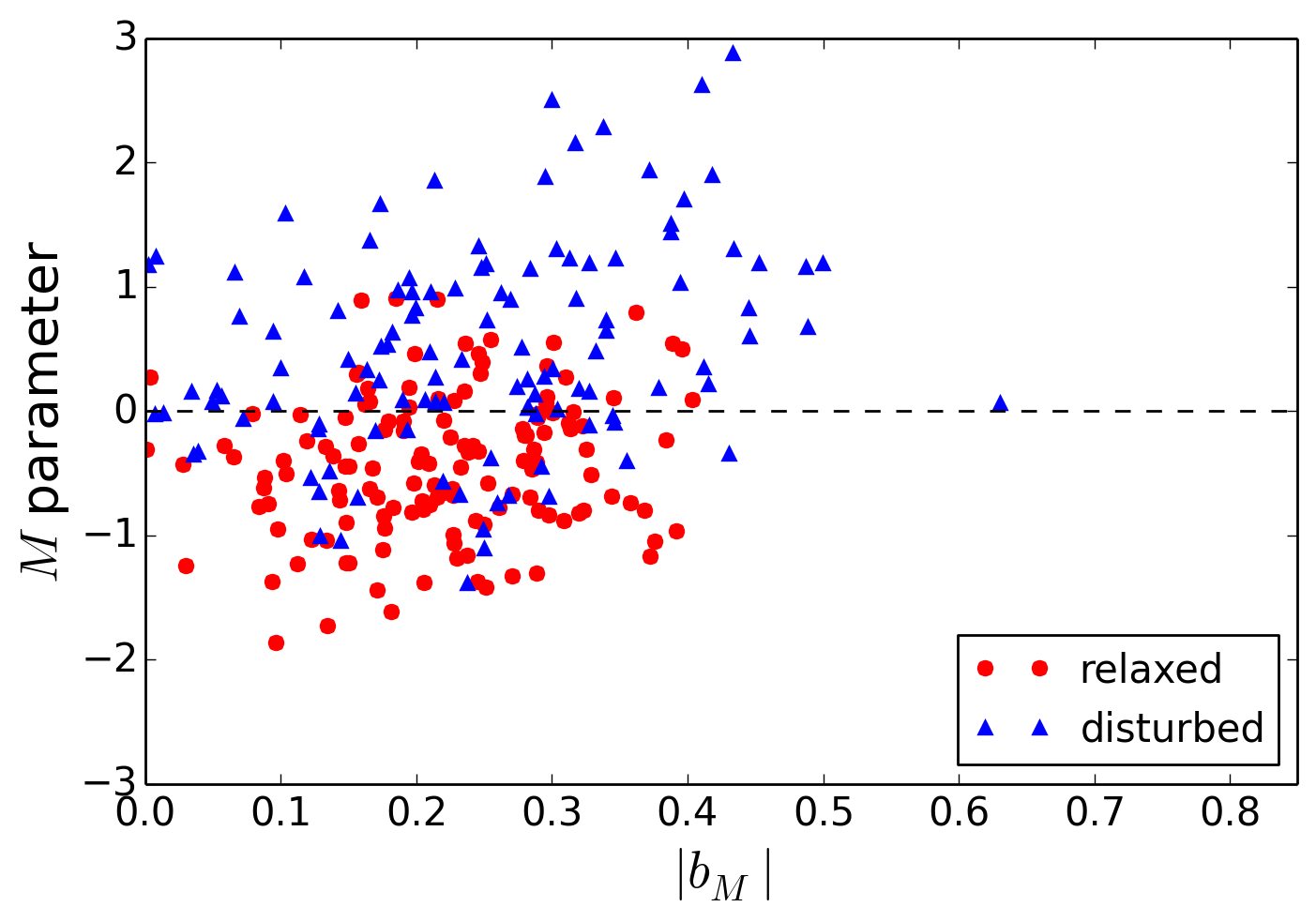}}
	\caption{Mass bias absolute value $|b_M|$ plotted against the $M$ combined parameter for the CSF
		flavour at redshifts 0.43, 0.54, 0.67 and 0.82 (shown in panels (a), (b), (c) and (d), respectively).
		Relaxed/disturbed clusters are marked with red filled circles/blue filled triangles.
		The black dashed line marks the threshold ($M=0$) for the $M$ parameter.}
	\label{fig: mass bias versus m parameter CSF}   
\end{figure*} 
\begin{table}
	\centering
	\caption{Pearson correlation coefficients (CC) found between $|b_M|$ and $M$ for all considered redshifts and flavours.}
	\label{tab:mass bias vs M, Pearson CC}
	\begin{tabular}{lcc} % three columns, alignment for each
		\hline
		$z$ &  \multicolumn{2}{c}{CC} \\
		 & CSF & NR \\
		\hline
		0.43 & 0.29 & 0.28 \\
		0.54 & 0.27 & 0.29 \\
		0.67 &  0.26 & 0.30 \\
		0.82 & 0.27 & 0.37 \\
		\hline
	\end{tabular}
\end{table}
\section{Morphology and projected SZ-CM offset}
\label{sec: projected shift correlation}

An additional indicator of the dynamical state which can be computed from observations, is the offset between the positions 
of the brightest cluster galaxy (BCG) and the X-ray peak.
The goodness of this parameter to infer the dynamical state has been proven through the years by both observations
\citep[see e.g.][]{Katayama_2003_BCG_X_shift,Patel_2006_BCG_X_offset,Donahue_et_al_2016,Rossetti_2016_BCG_offset}
and simulations \citep[see e.g.][]{Skibba_2011_BCG_X_offset_from_simulations}.
An equivalent indicator on SZ maps has been recently investigated by 
\citet{Gupta_et_al_2016:spatial_shift_arXiv161205266G} using the \emph{Magneticum} simulation. Indeed, in the aforementioned work they computed the 
projected offset between the centre of gravitational potential of the cluster and the SZ peak, normalised to $R_{500}$.

\begin{figure*}
	\centering
	\subfloat[]
	{\includegraphics[width=1\columnwidth]{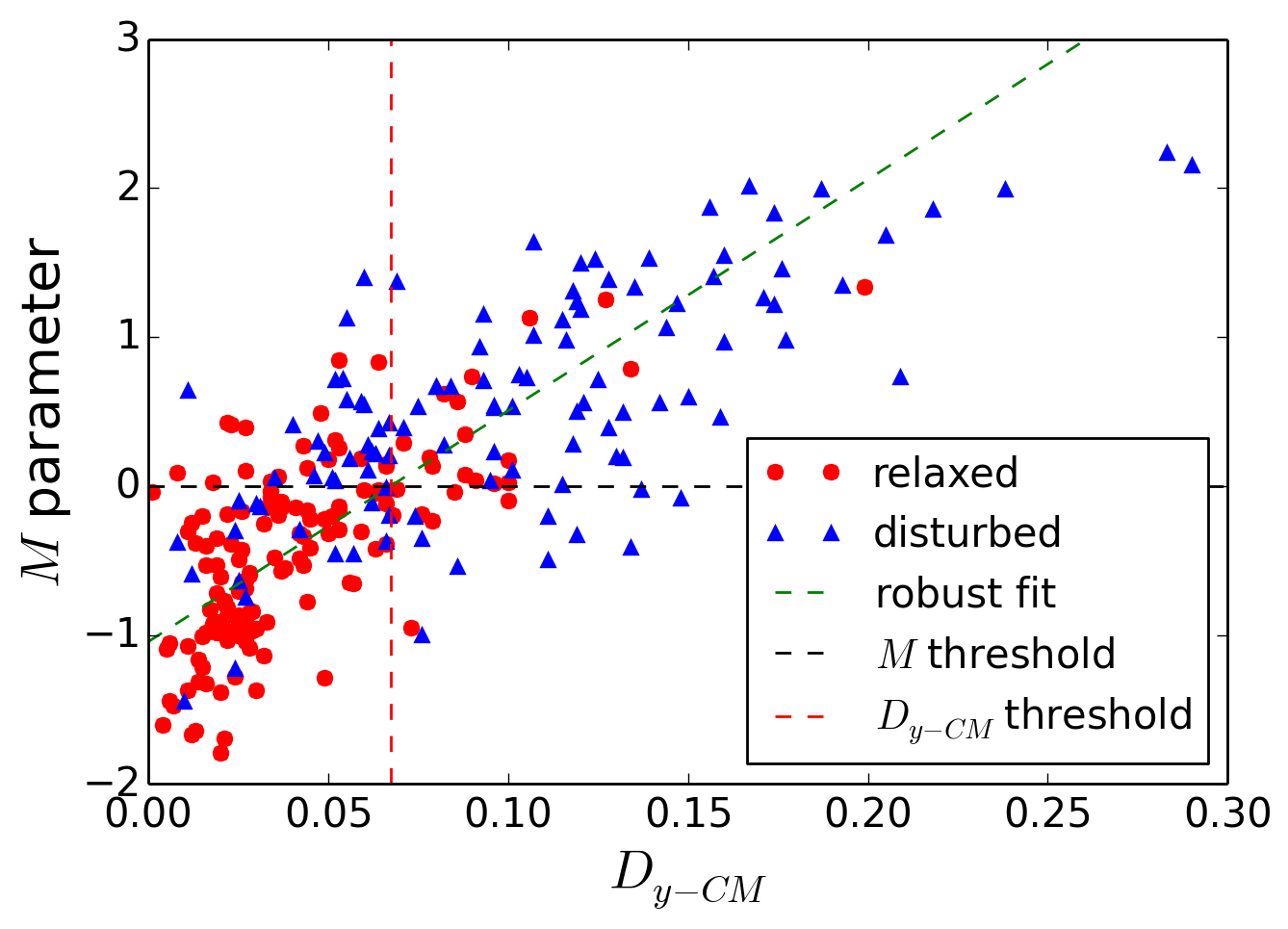}}
	\hspace{5mm}
	\subfloat[]
	{\includegraphics[width=1\columnwidth]{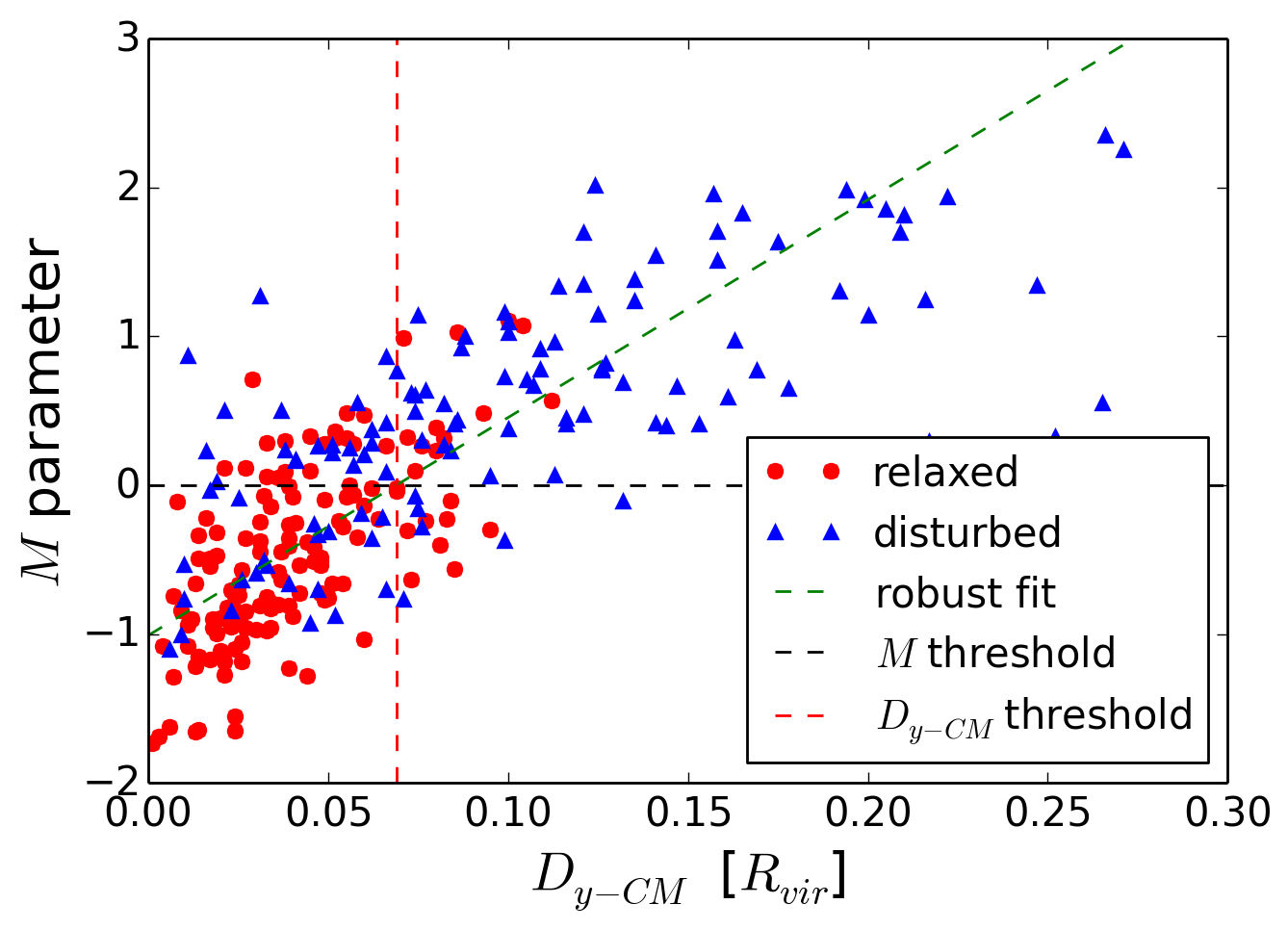}}
	\hspace{5mm}
	\subfloat[]
	{\includegraphics[width=1\columnwidth]{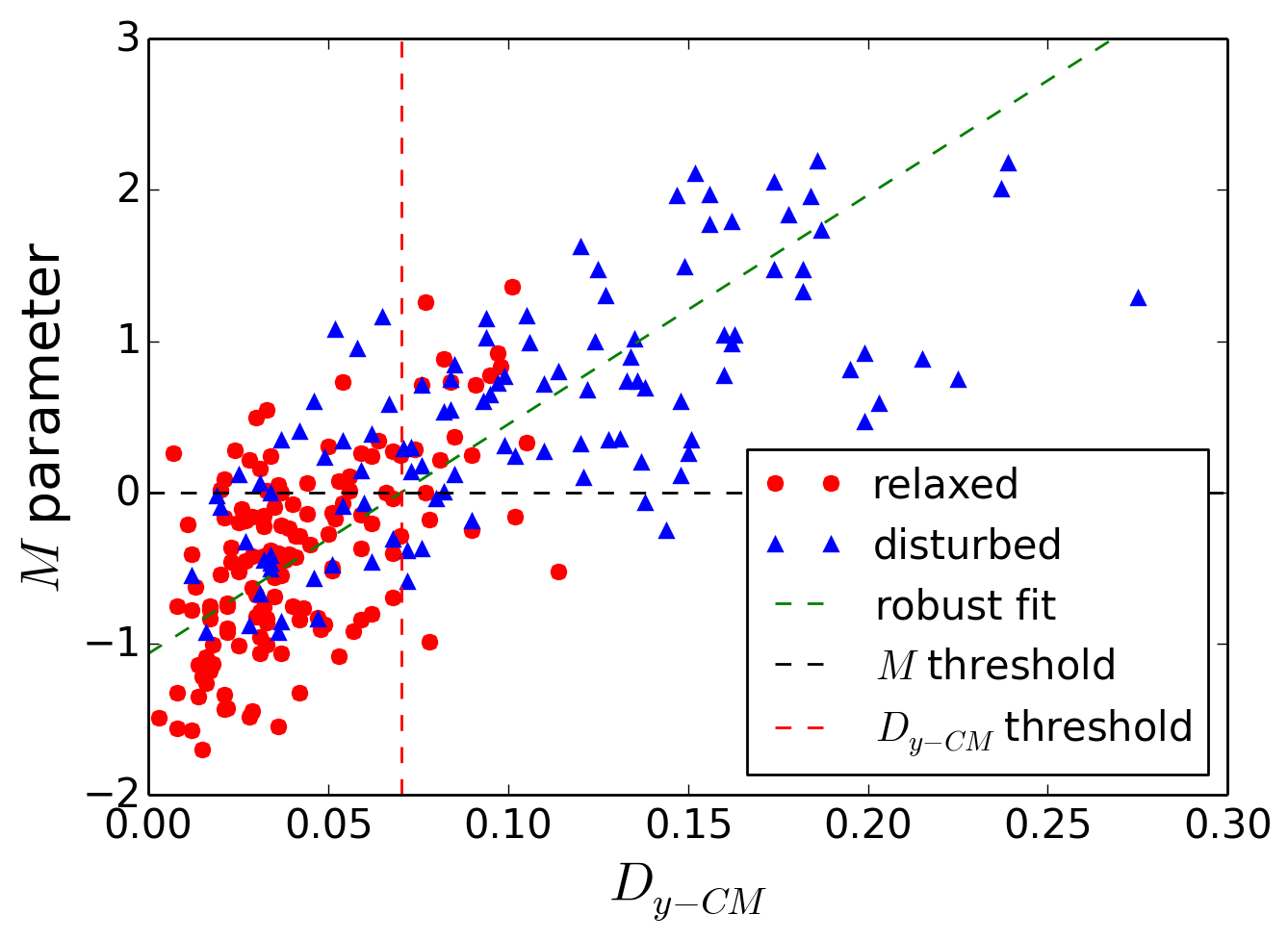}}
	\hspace{5mm}
	\subfloat[]
	{\includegraphics[width=1\columnwidth]{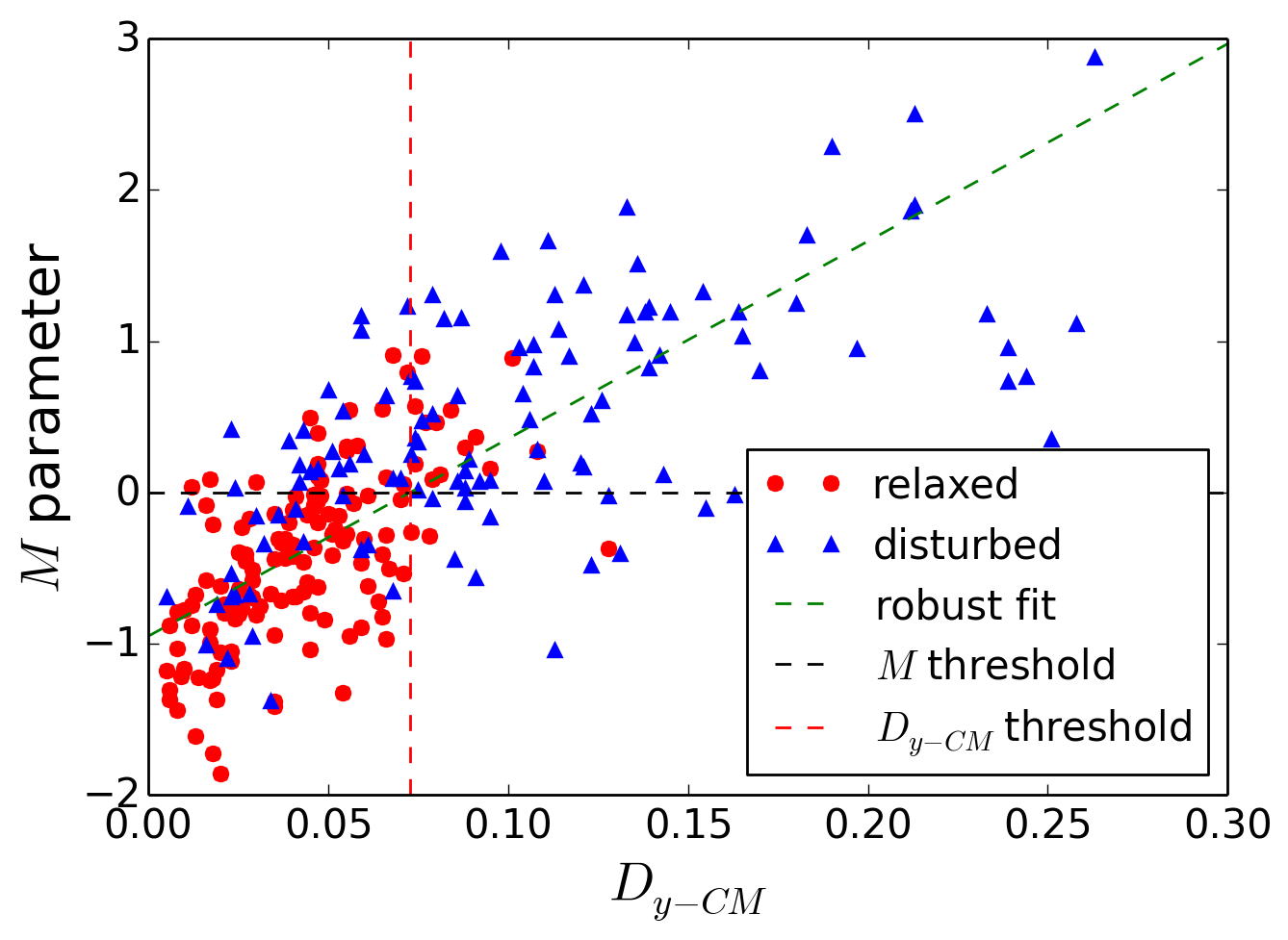}}
	\caption{Scatter plots of the $D_{y-CM}$ 2D offset vs the combined parameter $M$.
			Panels (a), (b), (c) and (d) refer to redshifts 0.43, 0.54, 0.67 and 0.82, respectively.
			Black horizontal and red vertical dashed lines mark the $M$ and $D_{y-CM}$ parameters thresholds, respectively. Green dashed lines show the best robust fit for all the clusters.}
	\label{fig: delta ctd versus m parameter CSF}   
\end{figure*}

We compute the projected offset $D_{y-CM}$ between the centroid of the $y$-map inside $R_{vir}$ and the 
centre of mass of the cluster 
%\textcolor{red}{%
(that at first approximation coincides with the BCG),
%}
normalised to $R_{vir}$. The correlation between $D_{y-CM}$ and our combined morphological parameter is then analysed, in order
to compare pure morphological information with an observational dynamical state-driven quantity. The correlation is of 
the order of 80\%, as shown in Table \ref{tab:delta ctd vs M, Pearson CC}. In Fig. \ref{fig: delta ctd versus m parameter CSF} the correlation for the CSF 
sub-set is shown; no significant deviations are found in the NR sample. The 
strong correlation between $ M $ and $D_{y-CM}$ suggests that morphology is closely related to the dynamical state. Therefore, under 
the assumption of coincidence between CM and BCG positions, this parameter can be easily inferred from joint 
optical and SZ observations. Being the threshold value for the $M$ parameter known, we derive the corresponding average threshold value for $D_{y-CM}$ of 0.070 by interpolating on the best-fit curve. This allows us to infer that the cluster is 
relaxed(unrelaxed) if $D_{y-CM}$ is below(above) this value. Moreover this threshold value is consistent with the one found by 
\citet{Meneghetti_et_al_2014:The_MUSIC_of_Clash:Predictions_on_the_Concentration-Mass_Relation} referring to the 3D offset 
between the position of the CM and the minimum of the gravitational potential. Following the same approach used for the 
analysis of the $M$ parameter contaminants, we find that there is an average of 25\%(20\%) contaminant 
clusters in the relaxed(disturbed) population. Being this contamination lower than the corresponding one for the $M$ 
parameter, we conclude that this 2D offset performs better in discriminating the dynamical state of the clusters. 
Considering the averaged deviation of $M$, $< {\sigma_M} >=0.41$ we derive through an interpolation the 
corresponding deviation of $D_{y-CM}$ of $ < {\sigma_D} >=0.029$. Hence, as a final criterion, we consider a cluster as 
relaxed if $D_{y-CM}<0.041$ and disturbed if $D_{y-CM}>0.099$.
\begin{table}
	\centering
	\caption{Pearson correlation coefficients (CC) between $D_{y-CM}$ and the $M$ combined parameter for all the 
		analysed redshifts and flavours.}
	\label{tab:delta ctd vs M, Pearson CC}
	\begin{tabular}{lcc} % three columns, alignment for each
		\hline
		$z$ &  \multicolumn{2}{c}{CC} \\
		& CSF & NR \\
		\hline
		0.43 & 0.78 & 0.78 \\
		0.54 & 0.77 & 0.75 \\
		0.67 &  0.76 & 0.79 \\
		0.82 & 0.75 & 0.75 \\
		\hline
	\end{tabular}
\end{table}
\section{Conclusions}
\label{sec:conclusions}

The study of cluster morphology through several indicators allows to analyse large amounts of data from surveys in 
different spectral bands and will be crucial for the understanding of the structure formation scenario. This 
topic has been already studied in the X-ray band, and we approach it here for the first time using SZ maps, opening 
a new window on cluster morphology in the microwave band. To this purpose, we analyse the application of some
morphological parameters on the  synthetic SZ maps of $\sim260$ massive clusters extracted from the MUSIC-2 data set, 
taking non-radiative and radiative physical processes into account, and studying four different redshifts.
The clusters have been a priori classified as relaxed and disturbed using two standard 3D theoretical indicators: $\Delta_r$ and
$M_{sub}/M_{vir}$ related respectively to the offset between the peak of the density distribution and the centre of
mass, and to the mass ratio between the biggest sub-structure and the total halo.
We use a set of observational parameters derived from X-ray literature and two new ones (the Gaussian fit
and the strip parameters), testing their performances when applied on the SZ maps in terms of efficiency and stability.
All the parameters have been properly combined into a single morphological estimator $M$. The discriminating power of M, is found to be higher respect to the single parameters. Moreover we studied its possible 
correlations with the hydrostatic mass bias and with the projected offset between the position of the SZ peak and the 
position of the centre of mass of the cluster.

The results we achieved can be summarized as follows:
\begin{itemize}
\item Few morphological parameters (namely the light concentration, the
centroid shift and the asymmetry) are efficient when applied to SZ maps, as they are for the X-ray maps. This may be because of their sensitivity to the very central core region morphology, which has been already well probed both in SZ and in X-ray imaging.
Nevertheless, other parameters have been proven to be less efficient with respect to the application to
X-ray map, as e.g. the third-order power ratio and the fluctuation parameter,
which, in particular, shows an unexpected behaviour in our case;

\item The combined parameter $M$ shows less overlap of the two dynamical state populations with respect to any of the single parameters. Its threshold value $M=0$ properly discriminates between relaxed and disturbed dynamical states.
The contaminant percentages are 22 per cent for disturbed clusters, and 28 per cent for relaxed, respectively.
We have estimated the error for the parameter as the standard deviation, computed from the estimation of M on maps projected along many different line of sight, finding it to be $ \overline{\sigma_M} = 0.41 $;

\item By varying the angular resolution of the maps, involving a Gaussian smoothing, the overlap of the two populations
determined through the combined parameter shows an expected increase at the lowest resolution we have tested (5 
arcmin). Nevertheless, when the resolution is below or equal to 1 arcmin the  performances are improved. This may be because of the small fluctuation suppression on the maps due to the convolution with the beam resolution;

\item The $M$ parameter, which has been proven to be a good proxy of the dynamical state, shows a weak correlation with the mass bias, in agreement with other analyses found in literature;

\item The 2D spatial offset $D_{y-CM}$ between the SZ centroid and centre of mass of the cluster is strongly correlated 
with its morphology and dynamical state, with a correlation coefficient of $ \sim 80 $ per cent,
This correlation shows no significant dependence on the cluster redshift or on the simulation flavour.
We propose this indicator as a fast-to-compute observational estimator of the cluster
dynamical state.
From the threshold value of 0.070 -- inferred by interpolating on the best-fit relation linking $D_{y-CM}$ with $M$ --
we get a contamination of 25 and 20 per cent of the relaxed and disturbed samples, respectively. This reduced contamination suggests a slightly better segregating power of this parameter as compared with $ M $. As a final criterion to distinguish between the two populations, we conclude that a cluster having
$D_{y-CM} < 0.041$ can be classified as relaxed, while it can be classified as non-relaxed if $D_{y-CM} > 0.099$.
\end{itemize}

%\textcolor{red}{%
We plan to apply the analysis outlined in this work to more realistic data, namely by processing the
current synthetic maps through the NIKA-2 instrument pipeline, in order to take the
impacts of astrophysical and instrumental contaminants into account.
%We also plan to apply this analysis on a new, larger set of simulated clusters. Given the remarkable growth in quantity and quality of SZ observations, that are expected to increase even more in the nearby future, the possibility of complementing the study of cluster morphology in X-rays through observations in the microwave band is a valuable tool for the characterization of such complex systems as galaxy clusters.
%}%

\section*{Acknowledgements}
%\textcolor{red}{%
The authors wish to thank the referee, G. B. Poole, for the constructive and useful comments which improved this paper significantly, 
%}% 
and they acknowledge V. Biffi for useful discussions.
This work has been partially supported by funding from Sapienza University 
of Rome - Progetti di Ricerca Anno 2015 prot. C26A15LXNR. The MUSIC simulations have been performed in the 
Marenostrum supercomputer at the Barcelona Supercomputing Centre, thanks to the computing time awarded by Red 
Espa\~{n}ola de Supercomputaci\'{o}n. GY and FS acknowledge financial support from MINECO/FEDER  under research 
grant  AYA2015-63810-P. ER acknowledge financial contribution from the agreement ASI-INAF n 2017-14-H.0.

%%%%%%%%%%%%%%%%%%%%%%%%%%%%%%%%%%%%%%%%%%%%%%%%%%



\bibliographystyle{mnras}
\bibliography{bibliography.bib} 

%%%%%%%%%%%%%%%%%%%%%%%%%%%%%%%%%%%%%%%%%%%%%%%%%%

%%%%%%%%%%%%%%%%% APPENDICES %%%%%%%%%%%%%%%%%%%%%

%\appendix



%%%%%%%%%%%%%%%%%%%%%%%%%%%%%%%%%%%%%%%%%%%%%%%%%%

% Don't change these lines
\bsp	% typesetting comment
\label{lastpage}
\end{document}